# Virial Clumps in Central Molecular Zone Clouds

*short title:* CMZ Virial Clumps

Philip C. Myers
Center for Astrophysics | Harvard and Smithsonian (CfA), Cambridge, MA 02138, USA
pmyers@cfa.harvard.edu

H Perry Hatchfield
University of Connecticut, Department of Physics, Storrs, CT 06269 USA

Cara Battersby
University of Connecticut, Department of Physics, Storrs, CT 06269 USA

## Abstract

CMZoom survey observations with the Submillimeter Array are analyzed to describe the virial equilibrium (VE) and star-forming potential of 755 clumps in 22 clouds in the Central Molecular Zone (CMZ) of the Milky Way. In each cloud, nearly all clumps follow the column-density-mass trend $N \propto M^s$, where $s = 0.38 \pm 0.03$ is near the pressure-bounded limit $s_p = 1/3$. This trend is expected when gravitationally unbound clumps in VE have similar velocity dispersion and external pressure. Nine of these clouds also harbor one or two distinctly more massive clumps. These properties allow a VE model of bound and unbound clumps in each cloud, where the most massive clump has the VE critical mass. These models indicate that 213 clumps have velocity dispersion $1 - 2$ km s$^{-1}$, mean external pressure $0.5 - 4 \times 10^8$ cm$^{-3}$ K, bound clump fraction 0.06, and typical virial parameter $\alpha = 4 - 15$. These mostly unbound clumps may be in VE with their turbulent cloud pressure, possibly driven by inflow from the galactic bar. In contrast, most Sgr B2 clumps are bound according to their associated sources and $N - M$ trends. When the CMZ clumps are combined into mass distributions, their typical power-law slope is analyzed with a stopped accretion model. It also indicates that most clumps are unbound and cannot grow significantly, due to their similar time scales of accretion and dispersal, ~0.2 Myr. Thus, virial and dynamical analyses of the most extensive clump census available indicate that star formation in the CMZ may be suppressed by a significant deficit of gravitationally bound clumps.

# 1. Introduction

## 1.1 Star Formation in the CMZ

The Central Molecular Zone (CMZ) is a distinct environment within our Galaxy, within a radius of ~250 pc, with physical properties that are closely matched to high-redshift galaxies (Kruijssen & Longmore 2013). Yet, at a distance of only 8.2 kpc (Gravity Collaboration et al. 2018, 2019; Reid et al. 2019), it is close enough to resolve individual star-forming cores, bringing this cosmologically representative region into our figurative back yard. The CMZ has a rich history of star formation, and currently contains about 2% of the stellar mass in the Milky Way (hereafter MW; e.g. McMillan 2017). It harbors three massive star clusters with ages of 2-6 Myr (the Young Nuclear, Arches, and Quintuplet clusters; Lu et al. 2013; Clark et al. 2018a, 2018b), and the fossil of a massive outflow, the Fermi lobes, which may be linked with a CMZ starburst within the past 10 Myr (Bordoloi et al. 2017, Su et al. 2010).

Despite its rich star formation history, the present-day star formation rate (SFR) of the CMZ falls short of expectations by about an order of magnitude, in relation to its immense reservoir of dense gas (Longmore et al. 2013, Immer et al. 2012). Gas denser than ~ $10^4$ cm$^{-3}$ is considered a precursor of imminent star formation, with some supporting evidence that a "critical density" of this order must be exceeded for star formation to begin (e.g. Krumholz & McKee 2005, Lada et al. 2010, 2012, Heiderman et al. 2010). The total SFR of the CMZ is ~ 0.05-0.1 M$_\odot$yr$^{-1}$ or 3-6% of the total rate in the MW (Robitaille & Whitney 2010, Chomiuk & Povich 2011, Barnes et al. 2017). This rate is similar to the fraction of molecular gas in the Milky Way (3 x $10^7$ M$_\odot$ or 4%, Dahmen et al. 1998). However, the average gas density in the CMZ is significantly greater than the MW average (e.g. Güsten & Henkel 1983, Mills et al. 2018). In relation to its gas density, the SFR in the CMZ is about an order of magnitude lower than in other regions of the Galaxy with the same density (Longmore et al. 2013).

## 1.2. What Suppresses Star Formation in the CMZ?

Many of the physical properties of molecular gas thought to be important in the star formation process are different in the CMZ than in most of the MW, due to the unique CMZ environment. The Galactic bar may drive gas into the CMZ, exciting strong turbulent motions (Hatchfield et al. 2021a; Sormani & Barnes 2019, Kruijssen et al. 2014). The resulting line widths are about 5-10 times larger than in the Galactic disk (Shetty et al. 2012, Kauffmann et al. 2017,

Henshaw et al. 2019). Gas in the CMZ is also subject to stronger magnetic fields (Crutcher et al. 1996, Pillai et al. 2015). It has higher densities (Güsten & Henkel 1983, Mills et al. 2018), higher temperatures (Mills & Morris 2013, Ginsburg et al. 2016, Krieger et al. 2017), and elevated cosmic ray ionization rates (Oka et al. 2005, Harada et al. 2015) compared to typical MW values.

The low SFR may be due to the relatively high level of turbulent pressure in the CMZ (Rathborne et al. 2014, Federrath et al. 2016). This turbulence may unbind or disperse clumps before they can fragment and collapse to form protostellar clusters. The turbulent driving mechanisms may arise from gas flows, star formation feedback, or from CMZ tidal forces and orbital dynamics (Krumholz et al. 2017, Dale et al. 2019, Kruijssen et al. 2019). Increased turbulence is expected in the nuclei of barred galaxies, as gas flows inward from the bar to the CMZ (Sormani & Barnes 2019). In more than 40 barred galaxies, nuclear gas on 150 pc scales is observed to have greater velocity dispersion, turbulent pressure, and virial parameter than in galaxy disks or than in nuclei of unbarred galaxies (Sun et al. 2020).

An alternate explanation is that star formation in the CMZ is episodic, in cycles between periods of low and high SFR. These cycles may arise from instability due to the short dynamical timescale of the region and the constant inflow of bar gas (Krumholz & Kruijssen 2015, Torrey et al. 2017).

Each of these explanations of a quiescent CMZ must also explain the anomalous properties of Sgr B2, whose star-forming burst is ~ 0.7 Myr old based on the time since pericenter passage (Ginsburg et al. 2018, hereafter G18). Sgr B2 is therefore contemporary with the quiescent CMZ clouds, suggesting that a common temporal cycle of star formation cannot account for all the CMZ clouds. Instead some starburst clouds may arise by collision of gas flowing from the bar to the CMZ with already orbiting CMZ gas (Sormani et al. 2020). A more detailed review of the foregoing topics is in Bryant & Krabbe (2021).

This paper addresses these issues of CMZ star formation by analyzing the first unbiased and complete survey of the high column density gas in the CMZ on 0.1 pc scales in the submillimeter (CMZoom: Battersby et al. 2020, hereafter B20; Hatchfield et al. 2020, hereafter H20). The observations are analyzed cloud-by-cloud with a virial equilibrium model. It indicates that most CMZ clumps are gravitationally unbound and either pressure-confined or transient. In a more global analysis, CMZ clump mass distributions (CMDs) are shown to have power-law slopes consistent with a model of stopped accretion. In a typical CMD the time scale of clump dispersal is similar to its time scale of accretion, of order 0.1 Myr. Together these results suggest that star formation is suppressed in most CMZ clumps because they cannot gravitationally bind their turbulent motions, and because they are likely to disperse before they can gain significant mass.

### 1.3. Plan of This Paper

In this paper, Section 2 presents a method of virial analysis of clump column densities $N$ and masses $M$ in the $\log N - \log M$ plane, which allows comparison with virial models when velocity dispersion observations are not available. Section 3 describes CMZoom observational techniques and data reduction. It presents the observations in the form of $\log N - \log M$ diagrams with virial models superposed. Section 4 describes linear trends in the $\log N - \log M$ data. It presents analysis in terms of virial equilibrium (VE) models, whose predictions of velocity dispersion, bound clump fraction, and virial parameter agree with other observations. Section 5 estimates the bound clump fraction in Sgr B2. It relates the two types of VE found in the CMZ to the VE in other parts of the MW and in nearby galaxies. Section 6 shows CMZ clump mass distributions (CMDs) which differ in their bound clump fractions and in the slopes of their power-law tails. In models of stopped accretion, the typical CMD slope indicates unbound clumps with similar time scales of accretion and dispersal, ~0.1 Myr. Section 7 presents a summary and discussion, and Section 8 gives the conclusions of the paper.

## 2. Virial Equilibrium Models of Molecular Clouds
### 2.1. Standard and Log $N$ − Log $M$ VE Models

The virial equilibrium (VE) model used here applies to a uniform, self-gravitating, magnetized sphere with random internal motions in a medium of uniform external pressure. It is based on equation (11-26) in Spitzer (1978 (hereafter S78; see also Bertoldi & McKee 1992 (hereafter BM92), McKee 1999 (hereafter M99), and Field et al. 2011, hereafter F11). In recent studies where the external pressure may be important, VE has been called "pressure-bounded virial equilibrium" (PVE). At one extreme of PVE, a cloud is in "simple virial equilibrium" (SVE; F11) when the external pressure is negligible and the cloud is gravitationally bound. At the opposite extreme, a cloud is "pressure-confined" by the difference between its external and internal pressure when its self-gravity is negligible (M99). In each case the cloud magnetic field tends to oppose its self-gravity, following the formulations in S78. A gravitationally bound cloud may depart from PVE by gravitational collapse, due, e.g. to an increase in external pressure and/or reduced internal velocity dispersion. An unbound cloud may depart from PVE by dispersal, if its external pressure is sufficiently turbulent (M99), or if its kinetic energy is much larger than its self-gravitating energy (Schruba et al. 2019), or if its virial parameter $\alpha$ exceeds the SVE criterion for gravitational binding, $\alpha \geq 2$ (BM92; Uehara et al. 2019).

The PVE model is implemented here in terms of mean column density $N$ and mass $M$ after substituting for spherical radius $R = [M(\pi m N)^{-1}]^{1/2}$, where the mean particle mass is $m$=2.33 $m_{\rm H}$ (Kauffmann et al. 2008). This "$N - M$" form of PVE analysis compares clump values of $\log N$ and $\log M$ to PVE models in the $\log N - \log M$ plane. This comparison is useful for clumps which have observed values of $N$ and $M$ but which lack observed velocity dispersions $\sigma$. In the CMZoom data, 755 clumps in the "complete" catalog of H20 are analyzed here with $N - M$ PVE, while ~80 clumps in the "robust" catalog of H20 were analyzed with standard PVE (Walker et al. 2018, Callanan 2021). The complete sample was chosen for analysis in this paper rather than the robust sample, in order to provide a larger statistical sample.

The $N - M$ PVE analysis can identify clumps that have similar mean density, if they follow the trend $N \propto M^s$, $s \gtrsim s_p = 1/3$. Here $s_p$ is the exponent in the pressure-bound limit of negligible self-gravity. This property is consistent with unbound clumps having similar pressure and velocity dispersion, as discussed in Section 2.2. It is prevalent among CMZ clumps, as shown in Section 3.3. Thus analysis with $N - M$ PVE is a useful alternative to standard PVE for estimation of clump binding in CMZ clouds, if they have a high fraction of unbound clumps and a low fraction of observed velocity dispersions.

Since the $N - M$ form of PVE analysis has not been presented before, this section compares plots of the $N - M$ PVE model with the standard PVE model in Figure 1. The next section, section 2.2, summarizes the $N - M$ PVE model equations and properties used to interpret the data in Section 3.

Figure 1 shows log-log plots of *(a)* $N$ vs. $M$ and of *(b)* $\sigma^2/R$ vs. $N$ for a PVE model. The format in *(a)* is more useful for continuum imaging surveys as in H20, where relatively few velocity dispersions are available, and where the observables $N$ and $M$ can be plotted on a log-log graph to compare with $N - M$ PVE models. The format in *(b)* is more useful for spectral-line imaging surveys where $N, R,$ and $\sigma$ are observables. Then $\sigma^2/R$ and $N$ are plotted on a log-log graph to compare with standard PVE models (Keto & Myers 1986, Heyer et al. 2009, F11, Walker et al. 2018, Callanan 2021). Equations relating $\sigma^2, M,$ and $N$ are given in section 2.2.

In Figure 1 each plot shows two curves with different values of external pressure. Each curve has a "pressure branch" where the external pressure is more important than self-gravity, and a "gravity branch" where self-gravity is more important than external pressure. These branches meet at the critical point, where the equilibrium is critically stable. The curves show that for different external pressures, the pressure branches are displaced and parallel, while the gravity branches coincide in the limit of strong gravitational binding.

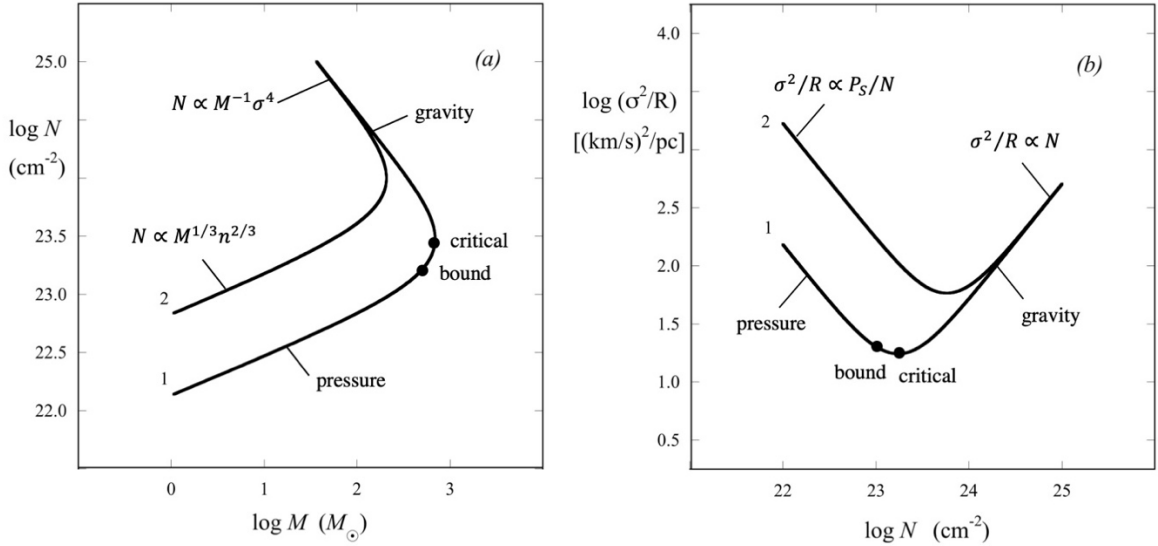

**Figure 1.** Pressure-bounded virial equilibrium (PVE) models *(a)* in $N - M$ form and *(b)* in standard form, for a uniform sphere of radius $R$ in a medium of constant external pressure $P_S$, for two different values of $P_S$. In each case the velocity dispersion is $\sigma = 2$ km s$^{-1}$ and the magnetic field strength is negligible. Curve 1 is the lower-pressure curve, with $P_S/k = 1.0 \times 10^8$ cm$^{-3}$ K, where $k$ is Boltzmann's constant. Its critical point has column density $N_0 = 3 \times 10^{23}$ cm$^{-2}$. Curve 2 is the higher-pressure curve, with $P_S/k = 9.2 \times 10^8$ cm$^{-3}$ K and $N_0 = 1 \times 10^{24}$ cm$^{-2}$. For each curve, the "bound" point marks the transition between the pressure-bounded and gravitationally-bound branches, where $\alpha/c_M = 2$. Each curve is labelled with the limiting power-law dependence of its *y*-variable on its *x*-variable. In $(a)$ the slope of the pressure-bounded branch approaches the limiting value $s_p = 1/3$, and the slope of the self-gravitating branch approaches $-1$. In $(b)$ the slope of the pressure-bounded branch approaches $-1$ and the slope of the self-gravitating branch approaches $1$. The $N - M$ curves resemble the standard curves after counter-clockwise rotation by $90°$.

Although the $N - M$ and standard versions of PVE analysis have many similarities, they are not equally useful alternatives. It is generally preferable to analyze the larger sample of useful information. Thus standard PVE analysis of line and continuum data has greater utility than $N -$

$M$ analysis of continuum data alone, when there are comparable measurements in lines and in the continuum. Only when the number of continuum measurements far exceeds the number of spectral line measurements, as in the present case, does $N - M$ analysis of the larger sample have potential to reach more useful conclusions than standard analysis of the smaller sample. In a more complete treatment, both samples should be analyzed together.

## 2.2. Log $N$ − Log $M$ PVE Properties

When the spherical radius $R$ is expressed in terms of $M$ and $N$, $R = [M(\pi m N)^{-1}]^{1/2}$, the virial equation (11-26) for a uniform sphere in S78 can be written as

$$M = (\pi m N)^{-1} \left(\frac{5\sigma^2}{G c_M}\right)^2 \left(1 + \frac{1}{3v^2}\right)^{-2} \tag{1}$$

where the magnetic factor is $c_M \equiv 1 - \lambda^{-2}$. Here $\lambda$ is the mass-to-flux ratio normalized by its critical value under ideal MHD, i.e. when the magnetic flux is strongly coupled to the gas mass. It is assumed that the CMZ clumps analyzed here are magnetically supercritical, $\lambda > 1$, so that their magnetic fields are not strong enough to prevent their gravitational contraction. This assumption is supported by Zeeman observations of nearby molecular clouds, which indicate numerous clouds with $\lambda > 1$, but no clouds with $\lambda < 1$ (Crutcher 2012).

If instead CMZ clumps were magnetically subcritical, $\lambda < 1$, they could contract along field lines on a free-fall time scale, but they could contract across field lines only via the process of ambipolar diffusion in nonideal MHD. These differing types of contraction would lead to flattened clump shapes over time scales much longer than the free-fall time. Significant observational evidence for these properties has so far not been reported (Crutcher 2012; Krumholz & Federrath 2019).

The CMZ clumps analyzed here are assumed for simplicity to have uniform internal density structure. This property is consistent with the conclusion in sections 3 and 4 that most observed clumps are pressure-bounded. If instead most clumps had centrally concentrated density as assumed in F11, the gravitational term in equation (1) would require multiplication by a factor of 1.2. The corresponding changes in clump mass and column density are relatively small compared to the scatter in the observational data.

The column density is normalized in the form $\nu \equiv N/N_0$, where the critical value for equilibrium is

$$N_0 \equiv \left(\frac{20 P_S}{\pi m^2 G c_M}\right)^{1/2}, \qquad (2)$$

and where $P_S$ is the external pressure on the surface of the sphere. Each of the curves in Figure 1 is a plot of equation (1) for $c_M = 1$ and for the values of $N_0$ and $\sigma$ specified in the figure caption. It should be noted that the critical column density $N_0$ is not a fundamental physical property, but it reflects the balance between surface pressure, gravity and magnetic field strength in critical virial equilibrium.

Useful special cases of equation (1) are described in the next three subsections.

**2.2.1. Pressure Binding** In the pressure-bounded limit $\nu \ll 1$, equations (1) and (2) yield

$$M \approx \left(\frac{9\pi m N^3}{16 n^2}\right)\left(\frac{P}{P_S}\right)^2, \qquad (3)$$

where $P$ is the mean internal clump pressure, $P_S$ is the external pressure on the surface of the sphere, and $n = P/(m\sigma^2)$ is the mean clump density. This limit is called "pressure-confined" by M99 or "pressure-bounded" by F11. In this limit, self-gravity is negligible and $P \approx P_S$.

For such a pressure-bounded PVE clump, the column density in equation (3) can be written

$$N = \left(\frac{16}{9\pi m^3}\right)^{1/3} n^{2/3} M^{1/3} \quad . \qquad (4)$$

If a group of such PVE clumps have similar velocity dispersions $\sigma$ and similar external pressures $P_S$ from clump to clump, their densities will follow $n \equiv P/(m\sigma^2) \approx P_S/(m\sigma^2)$, so $n$ will also be similar from clump to clump. Then equation (4) indicates that their column densities will display a correlation of the form $N \propto M^s$, $s \approx 1/3$ if the relative spread in their values of $n^{2/3}$ is small compared to the relative spread in their values of $M^{1/3}$. This property is used to interpret the observations in sections 3 and 4.

**2.2.2. Critical Gravitational Binding**  At the critical value $\nu = 1$, equation (1) gives the critical mass $M = M_0$, where

$$M_0 = \frac{9}{16\pi m N_0} \left(\frac{5\sigma^2}{G c_M}\right)^2 . \qquad (5)$$

Here $M_0$ is the maximum possible PVE mass for given values of $\sigma$, $P_S$, and $c_M$. Column densities greater than, equal to, or less than $N_0$ are respectively unstable, critically stable, or stable against gravitational collapse (S78).

In a $\log N - \log M$ plot equation (5) indicates that a group of clumps in a medium of constant pressure $P_S$ will lie along a line of slope $s = 0$ or constant column density $N = N_0$. Their mass $M = M_0$ will increase with velocity dispersion as $M \propto \sigma^4$. If instead they have constant velocity dispersion while $P_S$ varies, they will lie along a line of slope $s = -1$, where their column density $N = N_0$ increases with $P_S$ as $N \propto P_S^{1/2}$. If both $P_S$ and $\sigma$ vary from clump to clump, critically bound clumps on the $\log N - \log M$ plot will not follow a simple linear pattern. Thus an irregular pattern of points on a $\log N - \log M$ plot cannot be used to infer whether the clumps are critically bound, unless both $P_S$ and $\sigma$ are known.

**2.2.3. Strong Gravitational Binding**  As the column density becomes much larger than its critical value, $\nu \gg 1$, equation (1) becomes

$$M \approx \frac{1}{\pi m N} \left(\frac{5\sigma^2}{G c_M}\right)^2 \left(1 - \frac{2 P_S}{3 P}\right) , \qquad (6)$$

which approaches $M = (\pi m N)^{-1}[5\sigma^2/(G c_M)]^2$ in the limit of negligible external pressure $P_S$. If the magnetic field is also negligible, so that $c_M = 1$, this mass is the SVE "virial mass" (F11) and it is equal to $M = 5\sigma^2 R/G$ in terms of velocity dispersion and radius (BM92).

A group of such virial-mass clumps with $\nu \gg 1$ will lie along a line of slope $s = -1$ in the $\log N - \log M$ plot, where column density varies as $N \propto \sigma^4/M$ for constant $c_M$. However since they are gravitationally unstable against collapse and fragmentation, these clumps may not survive as long as gravitationally stable clumps having $3^{-1/2} \leq \nu \leq 1$.

In this case of strong gravitational binding, the trend of $N$ with $M$ in equation (6), $N \propto M^{-1}\sigma^4$, is much more sensitive to variations in $\sigma$ than in the opposite limit of pressure binding, in equation (4), when $N \propto M^{1/3}P_S^{2/3}\sigma^{-4/3}$. When velocity dispersions vary significantly from clump to clump, this difference may cause a trend of pressure-bound clumps to be better-defined than a trend of otherwise similar gravity-bound clumps, in the $\log N - \log M$ plane.

The foregoing relations indicate that the slope $d \log N / d \log M$ can be a useful diagnostic of the dynamical properties of a group of clumps in the $\log N - \log M$ plane.

**2.2.4. Virial Parameter and Log-Log Slope** When the PVE equation (1) is expressed in its usual terms of mass and radius, the virial parameter can be written $\alpha = c_M[1 - (P_S/P)]^{-1}$(McKee et al. 1993 eq. (34)). When the PVE equation is written in terms of mass and column density, $\alpha$ is obtained from its definition and from equation (1) in the equivalent form

$$\alpha = c_M\left(1 + \frac{1}{3v^2}\right) \quad . \tag{7}$$

The gravitational binding range is then $1 \leq \alpha/c_M \leq 2$, or equivalently $0 \leq v \leq 3^{-1/2}$.

The log-log slope $s \equiv d \log N / d \log M$ is a useful diagnostic of the binding of a group of clumps in the $\log N - \log M$ plane, as noted above. When $\sigma$ and $N_0$ are constant, differentiating equation (1) relates $s$ and $v$ by

$$s = \frac{(1/3)+v^2}{1-v^2} \quad . \tag{8}$$

A more general expression for $s$ is given in Section 5. The slope $s$ is related to the virial parameter by eliminating $v^2$ from equations (7) and (8), giving

$$s = \frac{1}{3-4c_M/\alpha} \quad . \tag{9}$$

Equation (9) shows that the slope $s$ approaches the limiting value $s_p = 1/3$ when $\alpha/c_M$ approaches $\infty$ on the pressure branch of the $\log N - \log M$ virial curve. This property is used to interpret the CMZ $\log N - \log M$ plots in Section 4.

Table 1 summarizes values of the normalized column density $v$, the normalized virial parameter $\alpha/c_M$, and the slope $s$ at key dynamical points along the PVE curve in Figure 1(a), based on equations (7) - (9) above.

**Table 1**

Key Parameter Values in $N - M$ Pressure-Bounded Virial Equilibrium

| (1) | (2) | (3) | (4) |
|---|---|---|---|
| PVE status | $v$ | $\alpha/c_M$ | $s$ |
| max PB | 0 | $\infty$ | 1/3 |
| bound | $3^{-1/2}$ | 2 | 1 |
| critical | 1 | 4/3 | $\infty$ |
| max GB | $\infty$ | 1 | -1 |

**Notes** - Column (1) lists key points along a pressure-bounded virial equilibrium (PVE) curve of $\log N$ vs. $\log M$ as in Figure 1(a), in order of increasing $N$. Here PB indicates the branch of pressure-bounded equilibrium, GB indicates the branch of gravitationally-bound equilibrium, and "bound" is the point between the two branches. Max PB is the extreme limit of PB, and max GB is the extreme limit of GB. The "critical" point is the turning point, where the mass has the greatest value allowed in equilibrium. Column (2) gives $v$, the column density normalized by its critical value. Column (3) gives $\alpha/c_M$, the virial parameter divided by the magnetic factor $c_M = 1 - \lambda^{-2}$. Here $\lambda$ is the mass-to-flux ratio normalized by its critical value. Column (4) gives the log-log slope $s = d \log N / d \log M$.

## 3. SMA Observations of CMZ Clouds

The Submillimeter Array[1] was used to carry out the CMZoom survey of the Central Molecular Zone (CMZ) of the Milky Way as a "Large Project" from May 2014 to July 2017. This survey was the first to give an unbiased census of submillimeter dust emission from the clouds of the CMZ, with FWHM resolution ~3 arcsec, or ~0.1 pc. The derived column densities range from $2.9 \times 10^{22}$ cm$^{-2}$ to $9.0 \times 10^{25}$ cm$^{-2}$, and the survey reports all the material in the CMZ with column density $N \gtrsim 10^{23}$ cm$^{-2}$ (B20). A detailed description of the observational results and a catalog of identified sources is given in H20.

---

[1] The Submillimeter Array is a joint project between the Smithsonian Astrophysical Observatory and the Academia Sinica Institute of Astronomy and Astrophysics, and is funded by the Smithsonian Institution and the Academia Sinica.

In this section, Section 3.1 gives a summary of the observational methods, Section 3.2 describes data and terminology, and Section 3.3 presents log $N$ - log $M$ plots of the observed data on a VE template for each of 24 clouds. Section 3.4 describes VE properties of the log $N$ - log $M$ plots, including nine which appear to describe virial systems of bound and unbound clumps.

### 3.1. Observational Methods, Data Reduction, and Cataloging Procedure

The CMZoom survey produced a complete map of all gas with $N \gtrsim 10^{23}$ cm$^{-2}$ in the central 5.0 deg longitude and 0.5 deg latitude of the Milky Way. The survey region was selected using *Herschel* column density maps (Battersby et al. 2011, Mills and Battersby 2017), including all gas with a column density $N(H_2) > 10^{23}$ cm$^{-2}$, excluding isolated bright pixels. The observations utilized the compact and subcompact configurations of the SMA, providing sensitivity to spatial scales between ~0.1 and ~2 pc at the distance of the Galactic Center. The data were calibrated according to standard SMA calibration procedures using MIR IDL. All imaging and deconvolution was performed in MIR IDL and CASA, and is described in detail in B20.

This work uses the CMZoom catalog of compact submillimeter emission presented in detail in H20. The sources characterized in this catalog are identified using a pruned dendrogram algorithm applied to the mosaic of CMZoom's 1.3 mm dust continuum emission. The dendrogram algorithm provides a tree-like hierarchical decomposition of the map's flux into discrete structures. The highest-level structures, called dendrogram 'leaves,' represent the brightest and most compact emission of the input map.

Due to the widely varying noise properties of the dust continuum emission detected by the SMA between different CMZ clouds, the initial dendrogram produced contains many false detections. To avoid these false detections, sources are removed from the source catalog unless they meet certain local noise conditions. Specifically, two catalogs are created: a "robust" catalog that prioritizes stricter pruning, and a "high completeness" catalog that prioritizes greater completeness. The high completeness catalog, used primarily in this work, is pruned such that all leaves included as sources in the catalog have a mean flux two sigma greater than the local noise, and a peak value four sigma greater than the local noise. Furthermore, the noise levels tend to be much higher near the edges of the surveyed region, and so leaves with their centroids within 10 pixels of the map's edge are flagged, but included in the high completeness catalog. Simulated interferometric observations of artificial maps with point-like sources were performed to determine the mass completeness of both catalogs. These tests indicate completeness greater than 95% for high-completeness sources more massive than 50 $M_\odot$, and for robust sources more massive than 80 $M_\odot$. These estimates are based on an assumed source dust temperature of 20K.

The mass of each source is determined in the isothermal approximation according to

$$M_{\text{leaf}} = \frac{d^2 S_\nu R_{dg}}{\kappa_\nu B_\nu(T_d)} \tag{10}$$

where $d$ is the galactic center distance, $S_\nu$ is the integrated source flux, $R_{dg}$ is the dust-to-gas ratio (~100, e.g. Battersby et al. 2011), $\kappa_\nu$ is the dust opacity per unit mass at frequency $\nu$, and $B_\nu(T_d)$ is the Planck function for dust temperature $T_d$ at frequency $\nu$. The column density of each leaf is found using the peak source flux density, $F_\nu^{\text{peak}}$ as

$$N_{H_2} = \frac{F_\nu^{\text{peak}} R_{dg}}{\mu_{H_2} m_H \kappa_\nu B_\nu(T_d)} \tag{11}$$

where $\mu_{H_2}$ is the mean atomic weight, taken to be 2.8 (Kauffmann et al. 2008), and $m_H$ is the mass of atomic hydrogen. The column density is based on the peak leaf flux, which is an average over the SMA beam, rather than the mean leaf flux. This choice is made because in some cases the dendrogram algorithm identifies leaves with a central core surrounded by a highly extended envelope. Nonetheless the column density is reduced by a factor 2.2 ± 0.3 based on values of mean flux and peak flux in Table 1 of H20. This reduction factor is consistent with the estimated systematic uncertainty in column density, as discussed in Section 7.2.2.

The uncertainties in the leaf mass and column density account for the local noise, typical variations in dust temperature, uncertainty in the distance to the Galactic Center, but are largely dominated in the systematic uncertainty in the dust opacity assumed, as the properties of dust grains are not well constrained in Galactic Center. Further details about the characterization of the catalog sources are presented in H20. Both the robust and high completeness catalogs, as well as all mosaics used to produce the catalogs have been made publicly available at https://dataverse.harvard.edu/dataverse/cmzoom.

## 3.2. Data and Terminology

This section displays data for column density $N$ and mass $M$ in the "high completeness" catalog in Table 5 of H20, for the 24 clouds having at least ten dendrogram leaves. These leaves have mass estimates in the approximate range $10\text{-}10^3\ M_\odot$. They are referred to here as "clumps" for consistency with the mass names of cores, clumps, and clouds in nearby star-forming regions (Bergin & Tafalla 2007, Table 1). It should be emphasized that many of the clumps in this catalog have further substructure visible with finer resolution, which may be more accurately described as "cores" (e.g. Lu et al. 2020). Nonetheless, the 755 clumps in this study provide the largest available sample of CMZ properties on the clump scale, for statistical comparisons of clumps within clouds, and from cloud to cloud.

## 3.3. Log $N$ − Log $M$ Plots for CMZoom Clouds

This section presents $\log N - \log M$ plots for the 24 high-completeness clouds in Table 5 of H20 having at least 10 dendrogram leaves, in four groups of six. Each cloud plot displays an average of ~31 clump data points, for a total of 755 clumps. Each cloud plot has error bars representative of the random error in $\log N$ and $\log M$ for the clumps in that cloud, according to H20 Table 6. Each variable is also subject to systematic errors of a factor of ~2, due mainly to uncertainty in assumed values of dust temperature and emissivity, as described in Section 3.1. These uncertainties shift all values of $\log N$ and $\log M$ by the same factor. Therefore systematic error bars are not shown, since the slope of a set of data points does not change if all points in the set shift in the same direction by the same amount.

Each plot contains a virial equilibrium template similar to the curves in Figure 1*(a)*, with velocity dispersions $\sigma = 1, 2, 4,$ and $8$ km s$^{-1}$, with critical column densities $\log N_0$ (cm$^{-2}$) $=$ 23, 24, and 25, and with negligible magnetic field strength, i.e. with $c_M = 1$. Each plot also contains a reference line of slope = 1, indicating clumps having an angular diameter of 2 arcsec or 0.08 pc. Since the synthesized beam of the CMZoom observations has diameter 3 arcsec (B20), no data point should lie to the left of the reference line, and barely-resolved clumps, if any, should appear close to this line.

Within each set of six cloud plots, the plots are ordered primarily according to increasing typical clump mass. The four sets (Figures 2a, 2b, 2c, and 2d) are also ordered by clump mass and by the presence of special features. Thus Figures 2a and 2b show clouds with the lowest-mass and next-lowest-mass clumps. Figure 2c shows clouds having low-mass clumps and one or two much more massive clumps. Figure 2d shows clouds having mostly moderate-mass clumps and also some much more massive clumps, culminating in G0.699-0.028 (Sgr B2), which has a uniquely broad distribution of low-mass, moderate-mass, and very massive clumps, including Sgr B2(M) and Sgr B2(N) (G18).

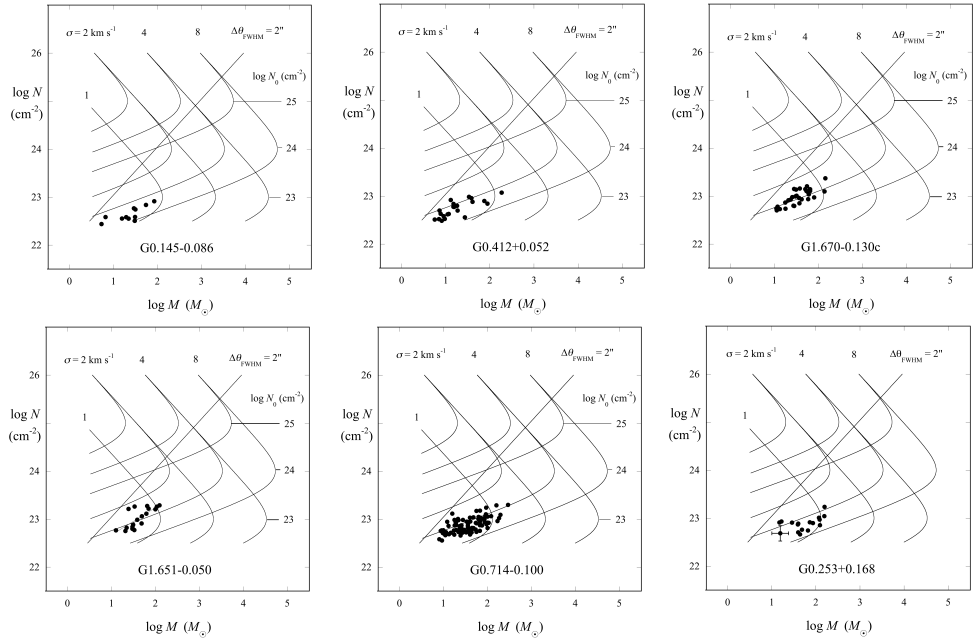

**Figure 2a.** Log-log plots of mean clump column density $N$ vs. clump mass $M$, for six CMZ clouds observed by the SMA CMZoom survey (B20, H20). The plots are displayed in order of increasing median $M$. The *line* indicates clumps with diameter 2 arcsec = 0.08 pc. *Curves* indicate virial equilibrium models of uniform spherical clumps as in Figure 1*(a)*, having velocity dispersion $\sigma = 1, 2, 4,$ and $8$ km s$^{-1}$, critical column density $\log N_0$ (cm$^{-2}$) = 23, 24, and 25, and magnetic factor $c_M = 1$. Each curve has a pressure-confined branch of limiting slope $s_p = 1/3$ and a self-gravitating branch of limiting slope -1. Most clumps follow a linear trend similar to the virial pressure-confined branch.

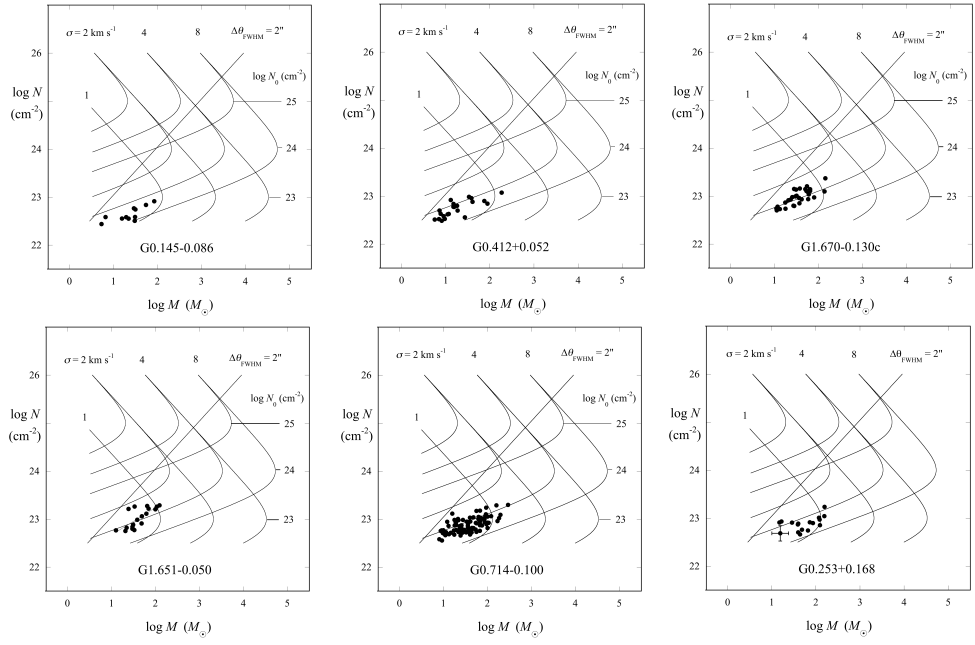

**Figure 2b.** Log-log plots of mean clump column density *N vs*. clump mass *M*, for six CMZ clouds as in Figure 2a, with slightly greater typical clump masses and with a similar linear trend.

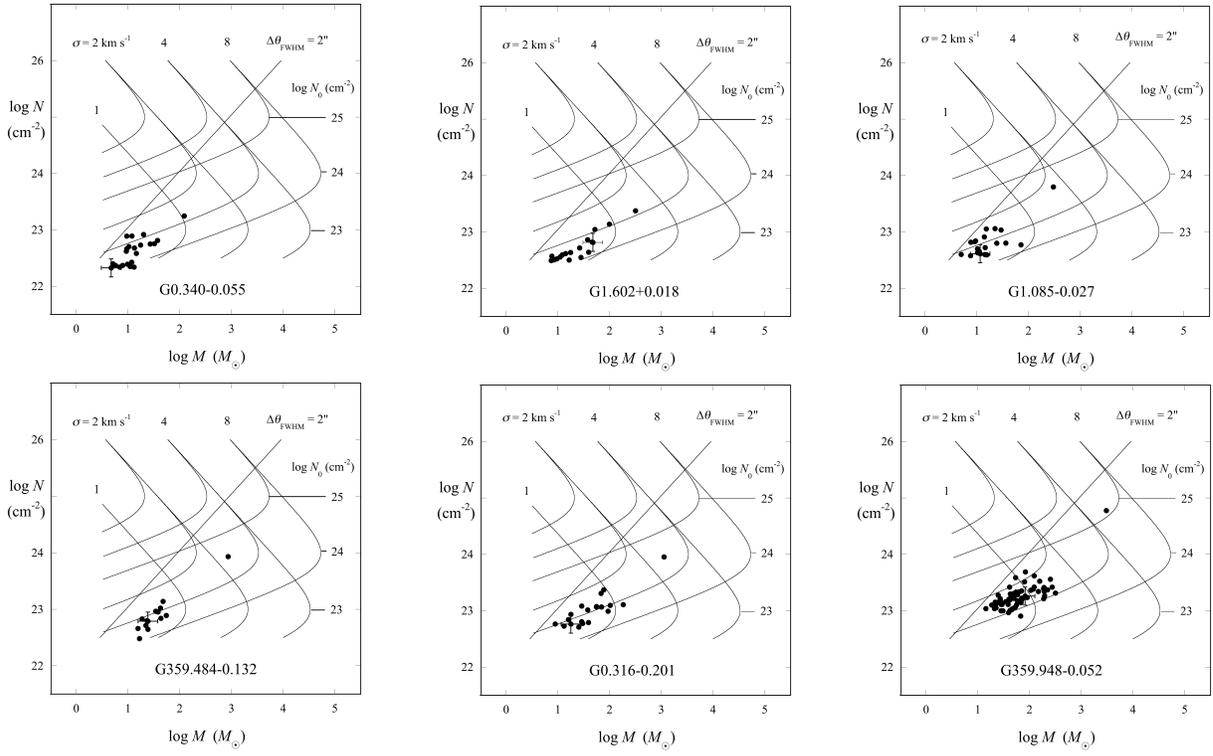

**Figure 2c.** log-log plots of mean clump column density $N$ vs. clump mass $M$, for six CMZ clouds as in Figure 2b, with slightly greater clump masses, with a linear trend, and each with one clump having much greater $M$ than all the rest.

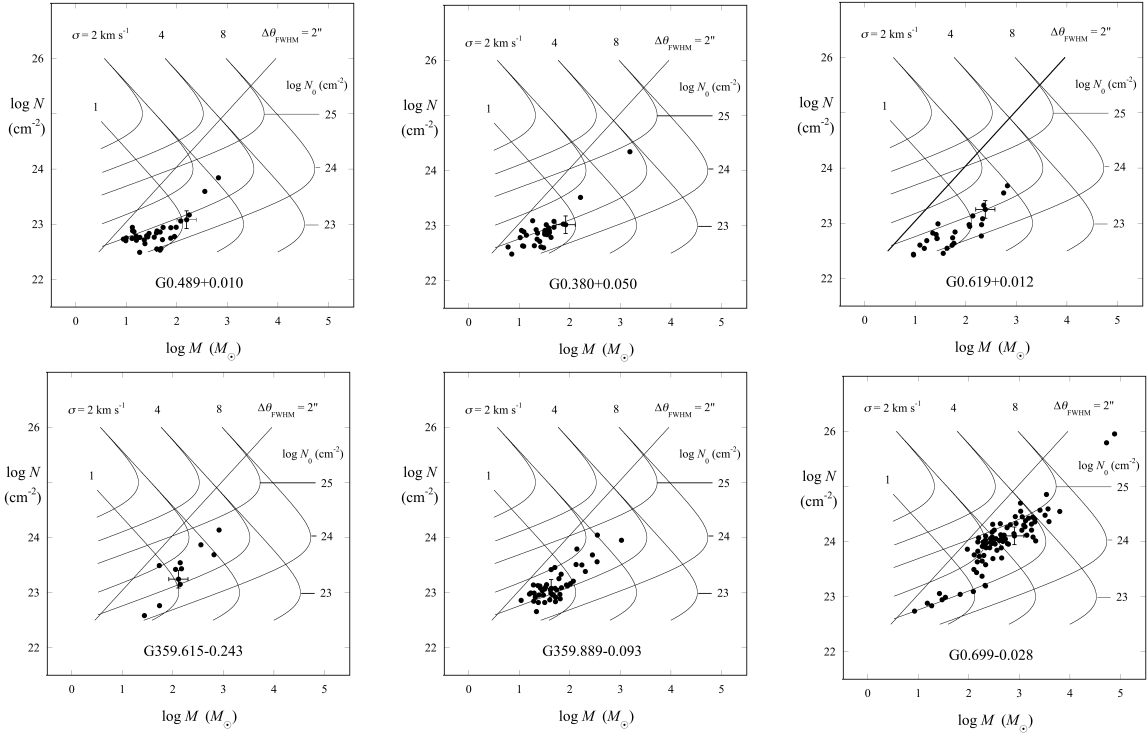

**Figure 2d.** log-log plots of *N vs. M* for six CMZ clouds as in Figure 2c, with a linear trend, or a linear trend and two more massive clumps, or with a more extensive trend of more massive clumps. The cloud with the most massive clumps is G0.699-0.028 (Sgr B2).

### 3.4. Log *N* − log *M* Clump Properties

In Figures 1*(a)*- 1*(d)*, two of the 24 clouds can be excluded from further analysis. The cloud G359.615 has been identified as a likely foreground object, not part of the CMZ, and the cloud G359.948 has been identified as having spurious image artifacts due to its close association with SgrA* (B20). The typical density, column density, mass, and estimated external pressure of the clumps in the remaining 22 clouds are given in Table 2.

## Table 2
### CMZ Cloud Properties

| (1) Cloud | (2) Clumps | (3) med $n$ ($10^5$cm$^{-3}$) | (4) med $N$ ($10^{22}$cm$^{-2}$) | (5) med $M$ ($M_\odot$) | (6) log ($P_{S0}/k$) (cm$^{-3}$K) |
|---|---|---|---|---|---|
| G0.001-0.058c | 57 | 1.7 | 6.0 | 26 | 7.4 |
| G0.068-0.075c | 26 | 1.4 | 5.0 | 21 | 7.5 |
| G0.070-0.035c | 13 | 1.1 | 4.7 | 23 | 6.9 |
| G0.106-0.082c | 15 | 1.3 | 3.5 | 11 | 7.4 |
| G0.145-0.086c | 11 | 1.1 | 3.8 | 29 | 6.9 |
| G0.253+0.016c | 17 | 1.6 | 7.9 | 70 | 7.5 |
| G0.316-0.201c | 20 | 2.7 | 10 | 38 | 9.0 |
| G0.326-0.085c | 25 | 1.0 | 3.2 | 13 | 7.0 |
| G0.340+0.055c | 22 | 1.5 | 4.4 | 11 | 7.6 |
| G0.380+0.050c | 32 | 2.2 | 7.4 | 31 | 9.7 |
| G0.412+0.052c | 23 | 1.9 | 5.9 | 29 | 7.2 |
| G0.489+0.010c | 35 | 1.9 | 6.0 | 36 | 8.7 |
| G0.619+0.012c | 26 | 1.4 | 6.1 | 54 | 8.4 |
| G0.699-0.028c | 85 | 2.8 | 110 | 400 | 13 |
| G0.714-0.100c | 95 | 1.9 | 7.4 | 36 | 7.7 |
| G1.085-0.027c | 20 | 2.0 | 6.1 | 14 | 8.7 |
| G1.602+0.018c | 18 | 1.3 | 4.2 | 22 | 7.8 |
| G1.651-0.050c | 18 | 3.0 | 11 | 36 | 7.6 |
| G1.670-0.130c | 30 | 2.9 | 9.5 | 34 | 7.8 |
| G359.484-0.132c | 14 | 3.1 | 6.8 | 29 | 8.9 |
| G359.734+0.002c | 23 | 0.96 | 3.3 | 15 | 7.0 |
| G359.889-0.093c | 48 | 3.6 | 12 | 44 | 9.1 |

**Notes** – Column (1) gives the cloud name as in Table 5 of H20, for all CMZ clouds with at least 10 dendrogram leaves (clumps) with signal-to-noise ratio $\geq 4$. Column (2) gives the number of clumps in the cloud. Column (3) gives the sample median of the mean clump densities, where $n = (9\pi m/16)^{1/2} N^{3/2} M^{-1/2}$. Column (4) gives the median clump column density, and Column (5) gives the median clump mass. Column (6) gives the log of the external pressure, calculated from equations (2) and (9) assuming that the most massive clump in the cloud is critically bound and that $c_M = 1$.

## 4. Analysis of $\log N - \log M$ Properties

### 4.1. Linear Trends

After the above exclusions the 22 clouds in the CMZ sample have 755 clumps, for a mean of 34 clumps per cloud. In the $\log N - \log M$ plots in Figure 2, the most prominent observational property is that nearly all clumps are well resolved, since in every plot nearly all data points lie significantly below the resolution reference line. The most prominent physical property in Figure 2 is the distribution of data points in closely spaced, approximately linear features, in the mass range $1 \lesssim \log(M/M_\odot) \lesssim 2$. In many cases the linear trend is nearly parallel to a pressure branch of its neighboring template lines, indicating a slope $\approx 1/3$. One special case is G0.699-0.028 (Sgr B2) in Figure 2(d). Only a small fraction of its clumps, with $\log N (\text{cm}^{-2}) < 23.2$, have a similar linear trend of $\log N$ with $\log M$, while most of its more massive clumps have a trend with steeper slope, as discussed in Section 5.2.

While many CMZoom leaves contain unresolved substructure on smaller scales, these $\log N - \log M$ trends cannot be due to the presence of unresolved sources, since then they would have a slope $s \approx 1$ rather than the observed $s \gtrsim 1/3$. They cannot be due to a fixed sensitivity limit on column density $N$, since then they would have slope $s \approx 0$. They cannot be due to strong gravitational binding, since then their slope would be close to -1 according to section 2.2.3. They arise independent of whether a local flux density background is subtracted as in the background-subtracted masses in H20 section 4.3. They also arise independent of whether the data are restricted to the "robust" source catalog of H20 section 4, or restricted to the "high completeness" catalog in the Appendix of H20.

The $\log N - \log M$ trends are similar to those we have identified in the data catalogs of other clouds observed with other telescopes, including Orion A N, observed with the JCMT (Salji et al. 2015), and G0.001-0.058 observed with ALMA (Uehara et al. 2019).

## Table 3
Fit Parameters for the Linear Trend $\log N = a + s \log M$ in Each Cloud

| (1) | (2) | (3) | (4) | (5) |
|---|---|---|---|---|
| Cloud | $a$ | $\sigma_a$ | $s$ | $\sigma_s$ |
| G0.001-0.058c | 22.21 | 0.07 | 0.42 | 0.05 |
| G0.068-0.075c | 22.35 | 0.09 | 0.28 | 0.06 |
| G0.070-0.035c | 22.5 | 0.1 | 0.12 | 0.07 |
| G0.106-0.082c | 22.16 | 0.06 | 0.40 | 0.05 |
| G0.145-0.086c | 22.2 | 0.1 | 0.33 | 0.09 |
| G0.253+0.016c | 22.5 | 0.2 | 0.22 | 0.09 |
| G0.316-0.201c | 22.3 | 0.2 | 0.4 | 0.1 |
| G0.326-0.085c | 22.2 | 0.1 | 0.32 | 0.08 |
| G0.340+0.055c | 22.0 | 0.2 | 0.6 | 0.2 |
| G0.380+0.050c | 22.4 | 0.1 | 0.34 | 0.09 |
| G0.412+0.052c | 22.3 | 0.1 | 0.33 | 0.06 |
| G0.489+0.010c | 22.5 | 0.1 | 0.20 | 0.07 |
| G0.619+0.012c | 22.1 | 0.1 | 0.42 | 0.07 |
| G0.699-0.028c | 22.5 | 0.1 | 0.29 | 0.06 |
| G0.714-0.100c | 22.40 | 0.06 | 0.28 | 0.04 |
| G1.085-0.027c | 22.5 | 0.2 | 0.2 | 0.1 |
| G1.602+0.018c | 22.0 | 0.1 | 0.49 | 0.07 |
| G1.651-0.050c | 22.1 | 0.2 | 0.6 | 0.1 |
| G1.670-0.130c | 22.3 | 0.1 | 0.45 | 0.07 |
| G359.484-0.132c | 21.7 | 0.3 | 0.8 | 0.2 |
| G359.734+0.002c | 22.1 | 0.1 | 0.37 | 0.08 |
| G359.889-0.093c | 22.1 | 0.1 | 0.60 | 0.06 |

**Note**- This table gives parameters for the least-squares linear fit to the trend of clump points in each cloud listed in Column (1). High-mass outlying clumps, if present, were not included in the fit to avoid biasing the linear slope. In the equation $\log N = a + s \log M$, $N$ has units of $cm^{-2}$ and $M$ has units of $M_\odot$. Column (2) gives the intercept $a$, Column (3) gives its one-sigma fit uncertainty $\sigma_a$, Column (4) gives the "cloud slope" $s$, and Column (5) gives its one-sigma fit uncertainty $\sigma_s$, according to the Levenberg-Marquardt algorithm (Levenberg 1944; Marquardt 1963).

To quantify these trends, the $\log N - \log M$ data were fit with linear models. This section describes fitting to obtain the distribution of the 22 best-fit "cloud slopes." As a check against possible bias due to different numbers of clumps in each cloud, a distribution was also obtained of the 595 "clump slopes," each defined with respect to the linear-fit intercept for its cloud. More detailed fits of the full virial model in equation (1) were made for nine clouds, as described in section 4.2.

Linear fits of the model $\log N = a + s \log M$ were carried out for each of the 22 clouds in Table 2, to obtain best-fit values of the intercept $a$ and slope $s$ in each cloud. For each of the 12 clouds in Figures 2(a) and 2(b), where the clump points have similar spacing, all clump points in the cloud were included in the fit. Where one or more points have distinctly greater $N$ and $M$ than the group of lower-mass points, as in Figures 2(c) and 2(d), the higher-mass points were excluded to avoid biasing the slope estimate. The fit parameters are given in Table 3.

The cloud-by-cloud distribution of best-fit linear slopes in the 22-cloud sample is shown in Figure 3. The typical best-fit slope $s$ is slightly greater than $1/3$, with mean, median, and standard error respectively 0.38, 0.36, and 0.03. The virial parameter corresponding to the mean slope is $\alpha = 11$, following equation (9), assuming that the magnetic field is negligible ($\lambda \gg 1$, $c_M = 1$). If instead the mass-to-flux ratio has a similar value to that in star-forming dense cores (($\lambda = 2$, $c_M = 3/4$, Crutcher 2012, Myers & Basu 2021), the corresponding virial parameter is $\alpha = 8$. In each case the virial parameters lie outside the range of gravitational binding, i.e. $\alpha > 2$. Instead they are consistent with pressure confinement, according to the criteria in Section 2.3.

The distribution of the 22 cloud slopes in Figure 3 may be limited by small-sample statistics. Also, the assignment of equal weight to each cloud could give some clumps more weight than others, since the number of clumps per cloud varies from 11 to 95 as given in Table 2. Therefore as a check a second slope distribution was calculated, where each of the 595 clumps has equal weight.

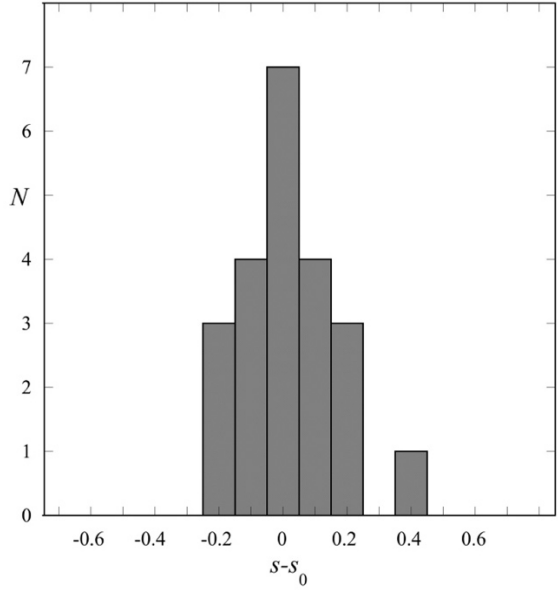

**Figure 3.** Distribution of log-log slopes $s$ for each of the 22 clouds in Figure 2, after exclusion of foreground and contaminated data. The display is centered on the reference value $s_0 = 0.38$. The linear fit parameters for each cloud are given in Table 3. In this distribution each cloud has equal weight. The typical slope values are mean ± standard error $0.38 \pm 0.03$ and median 0.36. These slopes are slightly greater than $s_p = 1/3$, the slope for the pressure-confined limit of $N - M$ PVE.

The simplest way to estimate the slope due to all the clumps would be to apply a linear fit to all 595 clumps in a single $\log N - \log M$ plot. However, according to the linear cloud fits in Table 3, the clumps in each cloud are associated with a different intercept as well as a different slope. Then a simple linear fit to the combined data would include both the variation of cloud slopes and the unwanted variation of cloud intercepts.

To reduce this bias a "clump slope" is defined as the slope of a line in the $\log N - \log M$ plane between a clump point $(\log M, \log N)$ and its cloud intercept point $(0, a)$,

$$s \equiv (\log N - a)/(\log M - 0) \quad . \qquad (12)$$

Here $a$ is the intercept in the linear fit through all the clump points in the cloud, given in Table 3. With this definition, each cloud slope in Table 3 is the mean of the clump slopes in that cloud. The resulting distribution of clump slopes $s$ for all 595 clumps in 22 clouds is shown in Figure 4.

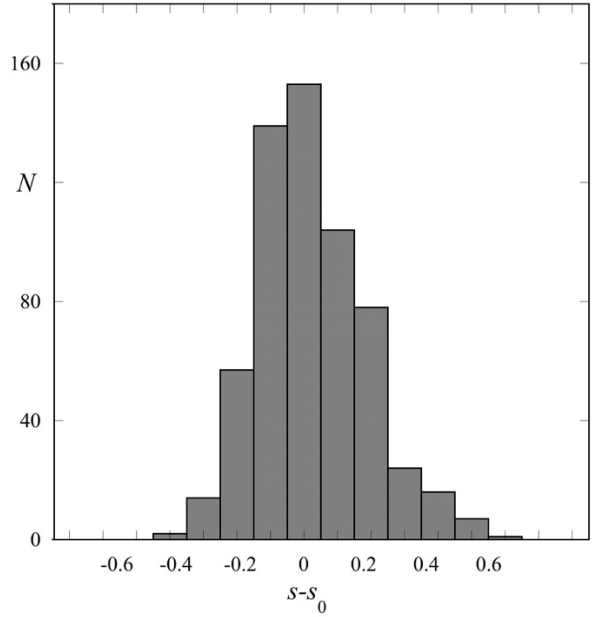

**Figure 4.** Distribution of log-log "clump slopes" $s$ defined in equation (12) for each of the 595 clumps in Figure 2, after exclusion of foreground and contaminated data. Each clump slope is the slope of a line from the clump point $(\log N, \log M)$ to its cloud intercept $(a, 0)$. The display is centered on the reference value $s_0 = 0.38$ as in Figure 3. The distribution of clump slopes has mean ± standard error = 0.38 ± 0.01 and median 0.36. These typical clump slopes are consistent with the typical cloud slopes in Figure 3.

The distribution of clump slopes in Figure 4 can be understood as the superposition of the clump slope distributions from each of the 22 clouds. Then the distribution in Figure 3 can be considered the result of replacing each of these 22 distributions by its mean value. The distribution in Figure 4 has mean, median, and standard error 0.38, 0.36, and 0.01. These values are consistent with the cloud slopes in Figure 3. This consistency indicates that the typical clump slope values in Figure 3 are not significantly compromised by statistical effects. Figures 3 and 4 each show a slight asymmetry favoring higher-slope values. This skewing may reflect the presence of a small proportion of bound clumps, discussed in Section 4.2.

In summary, the typical slope $s$ of a linear fit to clump points in the $\log N - \log M$ plane is slightly greater than the PVE limit of $s_p = 1/3$, by $s - s_p \approx 0.05 \pm 0.02$. This similarity of slopes suggests that most clumps in CMZ clouds are consistent with gravitationally unbound PVE, with the notable exception of G0.699-0.028 (Sgr B2).

A gravitationally unbound clump in PVE can maintain its identity if its confining external pressure is isotropic and time-steady, with relatively small fluctuations. This interpretation has been applied to high-latitude clouds (Keto & Myers 1986) and to dense cores in Orion (Kirk et al. 2017), where the external pressure is ascribed primarily to overlying cloud weight. On the other hand, if the external pressure is more turbulent, anisotropic and intermittent, with larger fluctuations, it may disperse the gravitationally unbound clump in a few crossing times. Then the clump is considered "transient" (M99, Uehara et al. 2019, Schruba et al. 2019).

### 4.2. Linear Trends with High-$M$ Outliers

In nine clouds in Figure 2, the clump points following the linear trend $\log N \propto s \log M$ are accompanied by one or two outlying points having significantly greater $M$, $N$, and $n$ than the other clumps in the cloud. These features are evident in clouds G0.068 in Figure 2*(a)*, G1.602, G0.340, G1.085 and G 359.484 in Figure 2*(c)*, and G0.619, G0.489, and G0.380 in Figure 2*(d)*. The linear trend in G0.699 (Sgr B2) is discussed separately in Section 5.1. In all nine of these clouds, the high-mass clumps are well-known star formation sites which have been studied previously.

These outlying high-mass points are not consistent with a simple extrapolation of the best-fit linear trend of the lower-mass points. Instead, each outlier has greater column density than the linear extrapolation. This change from constant slope to a steeper slope resembles the virial curve in Figure 1(a) near the "bound" point between pressure-confined and gravitationally bound gas. In that case, a virial curve which fits the $\log N - \log M$ data first passes through the unbound, pressure-confined points with slope slightly greater than 1/3, and then curves upward with increasing slope as it approaches the gravitationally bound points.

### 4.3. PVE Model Fits

The above suggestion that some $\log N - \log M$ plots resemble the virial curve of a "bound-and-unbound system" is tested here with PVE model fits. These fits were performed by eliminating $[(5\sigma^2)/(Gc_M)]^2$ from equations (1) and (4), yielding

$$M = \frac{16 M_0}{9}\left(\nu^{1/2} + \frac{1}{3\nu^{3/2}}\right)^{-2} \qquad (13)$$

where $\nu = N/N_0$. It is assumed that in each system, the maximum observed mass $M_{\max}$ equals the critical mass $M_0$, the greatest mass allowed in a virial system. Then equation (13) is used to make a least-squares fit of $\log M$ ($\log N$), which gives the best fit to the observed points and the best estimate of the critical column density $N_0$. The resulting nine fits are shown in Figure 5, and their fit parameters are summarized in Table 4.

In Figure 5, each fit curve passes through the middle of the low-mass points and passes near the most massive point. The single parameter in each fit, $N_0$, is determined within one-sigma relative uncertainty less than 10%. The value of $\nu$ for each data point is used to obtain the virial parameter $\alpha$ for each point, from equation (7), assuming $c_M = 1$. These estimates of $\alpha$ indicate that each massive point has $\alpha \approx 1$ and that the low-mass points have $2.5 \leq \alpha \leq 50$. The mean virial parameter is $\alpha = 14$ over all low-mass points and $\alpha = 10$ over all points. The corresponding mean log-log slope of the low-mass points is 0.40 from equation (8), similar to the value 0.38 found from the linear fits shown in Figures 3 and 4.

In summary, the results in Figure 5 and Table 4 are consistent with clumps in virial equilibrium with similar external pressure from clump to clump. In such a "bound-and-unbound system" one or two clumps have the greatest mass and the greatest column density in the cloud. They are gravitationally bound with virial parameter $\alpha < 2$, while ~20 lower-mass clumps are gravitationally unbound in PVE with mean $\alpha = 14$.

### 4.4. Comparison of Fit Results with Observations

The properties inferred from the PVE fits in Figure 5 apply to the nine clouds listed in Table 4. This subsection compares these derived quantities with available observations. The velocity dispersions are similar to those reported in two CMZ clouds observed with comparable resolution. The velocity dispersion $\sigma = 2.0$ km s$^{-1}$ in G0.489+0.010c (a.k.a. Dust Ridge e) is similar to ALMA line observations of $c$-C$_3$H$_2$, which give $\sigma = 1.1 - 2.1$ km s$^{-1}$ (Barnes et al. 2019). The dispersion $\sigma = 2.4$ km s$^{-1}$ in G359.484-0.132c (a.k.a. Sgr C) is similar to SMA observations of CH$_3$OH and N$_2$H$^+$, which give $\sigma = 1.0 - 2.3$ km s$^{-1}$ (Lu et al. 2019). These predicted values of $\sigma$ have been calculated by assuming that the magnetic field is negligible, i.e. the mass-to-flux ratio is $\lambda \gg 1$, so that $c_M = 1$. If instead the magnetic field is closer to its critical value, e.g. $\lambda = 1.2$, each of the predicted dispersions would be less by a factor of 1.8. Then each dispersion would more exactly match the mid-range of the observed values.

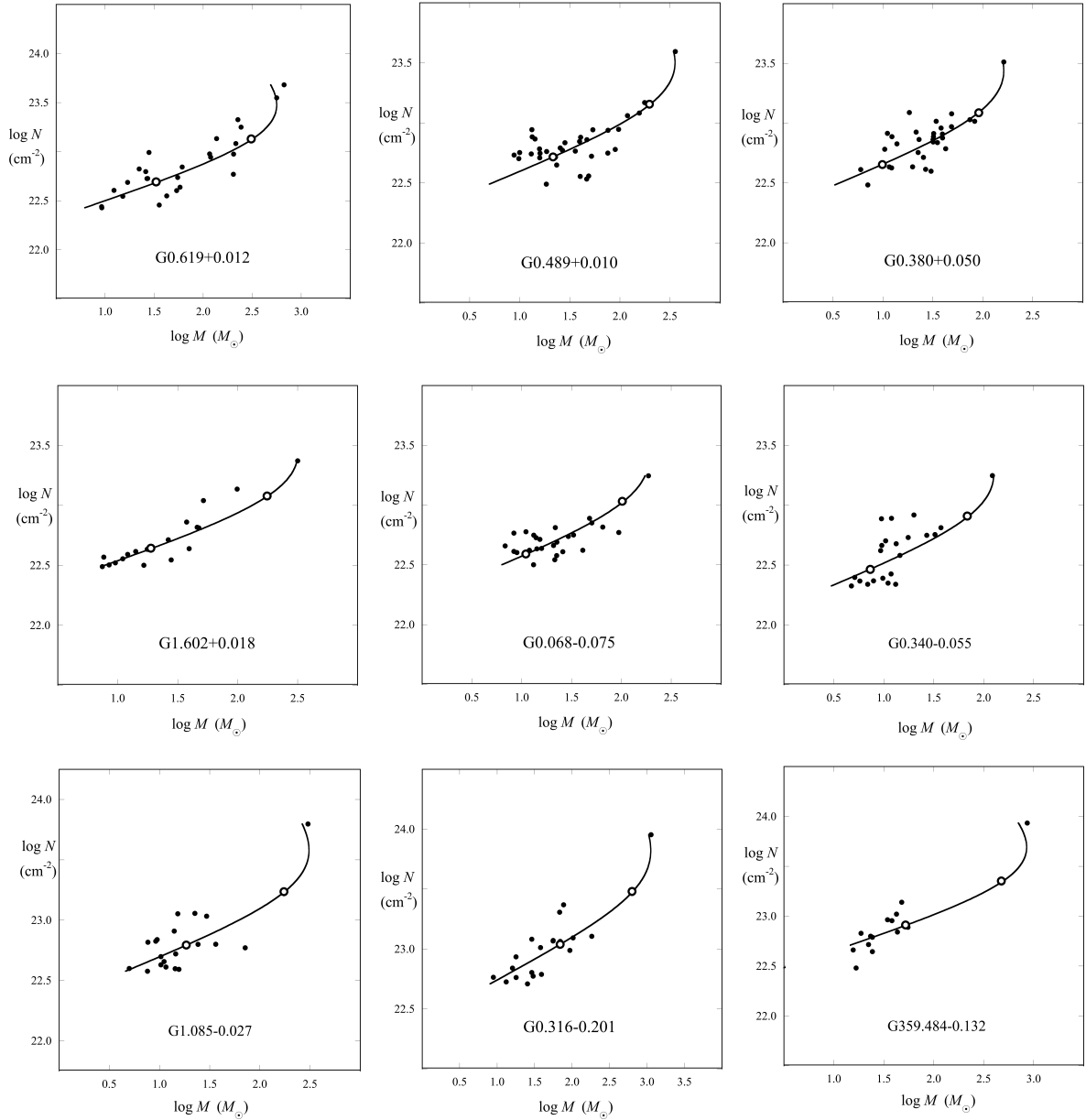

**Figure 5.** log *N* vs. log *M* for nine CMZ clouds whose clump population is consistent with one or two bound clumps (upper right) and ~20 associated unbound, pressure-confined clumps (lower left). The solid line is the best-fit virial model described in the text. Fit parameters are given in Table 4. Open circles indicate virial parameter $\alpha = 2$ (upper right) and $\alpha = 10$ (lower left). The $\alpha = 2$ position separates the gravitationally bound points from the unbound, pressure-confined points. The plots are displayed in order of increasing separation between the bound and unbound points. Estimated random errors are the same as shown in Figure 2.

**Table 4**

Virial Equilibrium Properties for Nine Bound-and-Unbound Systems

| (1) Cloud | (2) $N_0$ ($10^{23}$cm$^{-2}$) | (3) $M_0$ ($M_\odot$) | (4) $\sigma$ (km s$^{-1}$) | (5) $\alpha$ | (6) $f_b$ |
|---|---|---|---|---|---|
| G0.619+0.012c | 3.0 | 560 | 2.0 | 8.8 | 0.077 |
| G0.489+0.010c | 3.2 | 360 | 2.0 | 8.5 | 0.057 |
| G0.380+0.050c | 2.7 | 160 | 2.4 | 3.7 | 0.063 |
| G1.602+0.018c | 2.7 | 320 | 1.6 | 11 | 0.056 |
| G0.068-0.075c | 2.4 | 190 | 1.4 | 7.5 | 0.056 |
| G0.340+0.055c | 1.8 | 120 | 1.2 | 6.8 | 0.045 |
| G1.085-0.027c | 3.8 | 310 | 1.7 | 14 | 0.050 |
| G0.316-0.201c | 6.6 | 1100 | 2.8 | 15 | 0.050 |
| G359.484-0.132c | 4.9 | 850 | 2.4 | 17 | 0.071 |

**Note** – For each cloud in Column (1), Column (2) gives best-fit values of the critical column density $N_0$ based on fits of equation (9) to $M(N)$. The mean relative uncertainty $\sigma_{N_0}/N_0$ is 0.06. Column (3) gives the mass of the most massive clump, assumed to equal the critical mass $M_0$. Column (4) gives the velocity dispersion from equation (5), assuming $c_M = 1$. Column (5) gives the median virial parameter based on equations (1), (5) and (7), again assuming $c_M = 1$. Column (6) gives the bound fraction, or the ratio of the number of clumps with $\alpha < 2$ to the total number of clumps, given in Table 2. The clouds are listed in order of increasing separation between the bound and unbound points, as in Figure 5.

The identification of distinctly more massive, high-column-density clumps in Figure 5 and Table 4 as bound objects capable of forming stars leads to a bound fraction having mean ± standard error $f_b \approx 0.059 \pm 0.003$ in 213 clumps. This fraction is similar to several estimates of $f_b$ and related quantities in CMZ clouds. ALMA line observations of C$^{34}$S 2-1 and H$^{13}$CO$^+$ 1-0 in the 50 km/s cloud indicate $f_b \approx 0.04 - 0.07$ (Uehara et al. 2019). ALMA observations of three CMZ clouds including G359.484 show that 43 out of 834 continuum cores have associated SiO outflows, implying $f_b = 0.052$. These outflows indicate that the cores are protostellar, and therefore are mainly gravitationally bound (Lu et al. 2021). In a separate study, SMA observations of G359.484 based on virial parameters indicate $f_b = 0.09$ (Lu et al. 2019), similar to the estimate $f_b = 0.07$ made here.

Among bound clumps, the estimates of virial parameter $\alpha \lesssim 2$ in the clumps of maximum mass are similar to estimates of $\alpha$ in the maximum-mass cores in G0.489+0.010c, a.k.a Dust Cloud e (Barnes et al. 2019) and in the 50 km s$^{-1}$ cloud according to ALMA observations of the $J = 1$-$0$ line of H$^{13}$CO$^+$ (Uehara et al. 2019). Among unbound clumps, cores observed in the 50 km s$^{-1}$ cloud in the same line with $\alpha > 2$ have median $\alpha = 13$ (Uehara et al 2019). This virial parameter is similar to $\alpha = 10$, the mean of the clump virial parameters in Table 4.

In summary, the velocity dispersions, bound fractions, and virial parameters of the bound-and-unbound systems in Table 4 are similar to earlier estimates in a few individual CMZ clouds. This agreement justifies the assumption made above, that in each system the mass of the most-massive, highest-column-density clump is close to the PVE critical mass $M_0$, defined in equation (5).

## 5. PVE in Sgr B2 and in Galactic Environments

### 5.1. Association with ALMA Point Sources

The CMZ cloud G0.699-0.028 (Sgr B2) is remarkable among molecular clouds in the Milky Way in its high gas density and prolific star formation (G18; Schwörer et al. 2019). It is also remarkable compared to the 21 other clouds in this CMZ sample in its high clump masses and column densities (H20). This section estimates the fraction of gravitationally bound clumps in Sgr B2 based on comparison of the SMA observations at 1.3 mm with 3 arcsec resolution and the ALMA observations (G18) at 3 mm with 0.5 arcsec resolution. The next section 5.2 shows that most clumps in the $\log N - \log M$ diagram for Sgr B2 may also be consistent with PVE models of bound and unbound clumps.

The virial status of Sgr B2 clumps is estimated empirically by comparison between the SMA clumps in this paper and the ALMA sources in G18. It is first necessary to flag possible artifacts of compromised imaging near the strongest sources, as discussed in H20 and illustrated in H20 Figure 27. Accordingly SMA sources within ~1 arcmin of Sgr B2N and Sgr B2M which do not match positions of 3 mm point sources or H II regions were set aside, leaving 55 clumps for analysis.

These SMA clumps have a high degree of association with 3 mm ALMA sources. Of these 55 clumps, 43 have a 3 mm point source within 2 SMA beam diameters from the center position of a resolved SMA clump. Four clumps have a position match with strong HII emission, while eight clumps have no association with a 3 mm point source or with H II emission. This high degree of clump association with signposts of star formation in Sgr B2 is in sharp contrast with the 21 other clouds in the CMZoom sample (C21; Hatchfield et al. 2021b). Similarly, the Sgr B2 clumps which appear unbound according to their positions in the $\log N - \log M$ plot have no association with signposts of star formation in seven out of ten cases.

The 3 mm point sources probably require internal heating to be detected, and such sources are estimated to require typical luminosity $\sim 10^4\ L_\odot$ (G18). If so, they are probably unresolved high-mass protostellar objects (HMPOs) which are still accreting mass from their surrounding clump gas. If they resemble well-resolved HMPOs they also have accompanying clusters of lower-mass young stellar objects (G18), as observed in numerous young clusters. These properties suggest that at least $43/55 = 0.78$ of the SMA clumps are associated with gravitationally bound regions with ongoing star formation. As a conservative estimate the bound clump fraction $f_{b,\text{Sgr B2}} \approx 0.7 \pm 0.1$ is adopted.

Thus a much greater fraction of clumps in Sgr B2 are gravitationally bound than in the other 21 clouds in the CMZoom sample, ~0.7 as compared with ~0.06. A more detailed quantitative analysis can be made once well-resolved velocity dispersions become available for more clumps in Sgr B2 and in the other *CMZoom* clouds.

## 5.2. Model Fits in the $\log N - \log M$ Diagram

Most of the clumps in Sgr B2 may be consistent with gravitational binding, according to an analysis of $\log N - \log M$ slope similar to that of the unbound clumps in Section 3. Figure 6 reproduces the data points in the Sgr B2 plot in Figure 2(d) at a larger scale and without template curves, to show their properties more clearly. The data show two linear trends. Ten low-mass clumps have slope $s = 0.29 \pm 0.04$ with correlation coefficient 0.9, resembling the masses and slopes of unbound clumps discussed earlier in Section 3. These clumps are spatially distributed around the outskirts of the Sgr B2 region, where their surrounding gas has much lower density and pressure than in the filaments hosting the most massive clumps. In contrast, Figure 6 also shows that 73 more massive clumps follow an increasing trend in the range $23.3 < \log N < 25.0$. This trend is well-fit by a linear function with slope $s = 0.53 \pm 0.05$, with correlation coefficient 0.8.

To relate this steeper slope to bound clump models, the log of $M$ in equation (1) is differentiated with respect to $\log N$ as in Section 2.2.4. This more general derivation does not assume that the velocity dispersion $\sigma$ and the critical column density $N_0$ are constants. Then the resulting expression for the slope $s \equiv d \log N / d \log M$ is

$$s = \left[ 4 \frac{d \log \sigma}{d \log N} + \left( \frac{4}{1+3v^2} \right) \left( 1 - \frac{d \log N_0}{d \log N} \right) - 1 \right]^{-1} \tag{14}$$

where all quantities have the same definition as in Section 2. As expected, equation (14) reverts to equation (8) when $d \log \sigma / d \log N = d \log N_0 / d \log N = 0$.

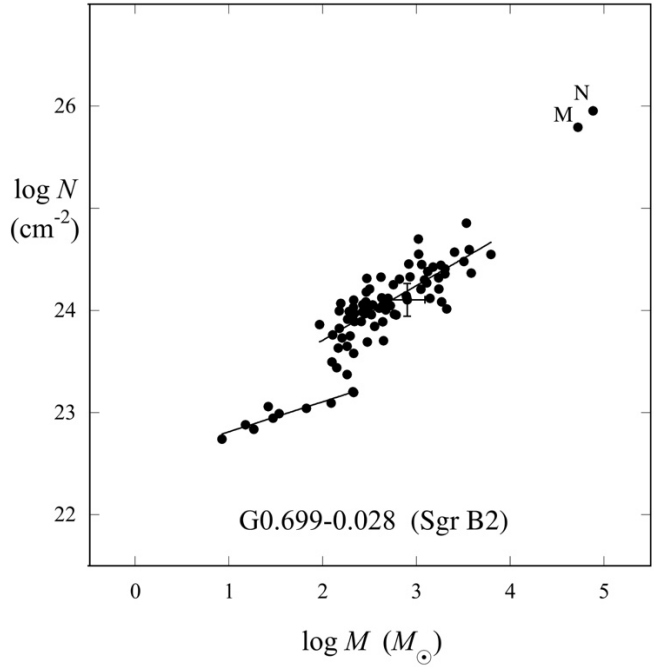

**Figure 6.** $\log N - \log M$ plot of clumps in G0.699-0.028 (Sgr B2). The labels M and N indicate the massive H II region clusters Sgr B2 (Main) and Sgr B2 (North). This plot has the same data as in Figure 2d, with best-fit lines added and virial model template lines removed. The lowest-mass clumps are fit with a line of slope $0.29 \pm 0.04$ and correlation coefficient 0.9. Their masses and slope resemble those of the unbound clumps discussed in Section 3. The 73 more massive Sgr B2 clumps have a best-fit line of slope $0.53 \pm 0.05$ and correlation coefficient 0.8. This slope

is fit by a PVE model of critically bound clumps, whose velocity dispersions increase with clump mass.

Equation (14) gives a condition on the velocity dispersion $\sigma$ for consistency with critically stable clumps following $s = 0.53 \pm 0.05$, which is approximated here as $s \approx 1/2$. Since critically stable clumps have $N = N_0$ or equivalently $\nu = 1$, equation (14) requires $d \log \sigma / d \log N \approx 3/4$, or equivalently $d \log \sigma / d \log M \approx 3/8$. This approximate value of $3/8$ is consistent with propagation of the fit uncertainty for $s$, which gives $d \log \sigma / d \log M = 0.38 \pm 0.01$. Substituting from the definitions of $N, N_0$, and $R$ gives the associated set of power-law relations between clump properties and $M$ as $N \propto M^{1/2}, \sigma \propto M^{3/8}$, $\rho \propto M^{1/4}$, $R \propto M^{1/4}$, and $P_S \propto M$.

These predicted relations should be tested against observations of bound clumps in Sgr B2 and other CMZ clouds. The trends $R \propto M^{1/4}$ and $\sigma \propto M^{3/8}$ are qualitatively plausible since more-massive star-forming clumps tend to be larger and more turbulent than their less-massive counterparts. A quantitative comparison can be made using observations of HCN J = 4-3 lines in numerous CMZ clouds. There, correlations of the form $M \propto R^{2.7 \pm 0.4}$ and $\sigma \propto R^{0.7 \pm 0.1}$ are reported, based on observations with a 20" beam (Tanaka et al. 2018). Combining these relations yields $\sigma \propto M^{0.3 \pm 0.2}$, which is statistically consistent with the predicted relation $\sigma \propto M^{3/8}$. A more detailed comparison should be made with finer-resolution observations, preferably limited to clumps in the Sgr B2 region.

In summary, most of the $\log N - \log M$ data for Sgr B2 show a power-law trend $N \propto M^{1/2}$, which is significantly steeper than the trend $N \propto M^p$, $p \approx 1/3$ for unbound clumps discussed in Section 2. Most of the Sgr B2 clumps are expected to be bound based on their close association with candidate high-mass protostars. The trend $N \propto M^{1/2}$ is consistent with a PVE model of critically bound clumps, where clump velocity dispersions increase approximately with clump mass to the 3/8 power. A preliminary comparison with line observations in several CMZ clouds shows consistency with this prediction, but more detailed comparisons are needed.

It appears that all 22 CMZ clouds may be described as virial systems of bound and unbound clumps, where 21 clouds have a low bound clump fraction while Sgr B2 has a much higher bound clump fraction.

### 5.3. PVE in Galactic Molecular Clouds

The bound and unbound types of PVE in CMZ clouds discussed above are similar to those seen in molecular clouds in the Milky Way (MW) and other nearby galaxies. A compilation of cloud properties in 12 regions in the MW and in seven nearby galaxies indicates that the clouds can be placed into two distinctly different PVE categories (Schruba et al. 2019 Tables 1 and 2). Unbound giant molecular clouds (GMCs) have low mean surface density $\Sigma_{GMC} = 10 - 30\ M_\odot\ pc^{-2}$ ($0.5 - 1.5 \times 10^{21} cm^{-2}$) and high virial parameter $\alpha = 3 - 10$. In contrast, bound clouds have high surface density $\Sigma_{GMC} = 10^2 - 10^4\ M_\odot\ pc^{-2}$ and low virial parameter $\alpha = 0.7 - 3$. The bound and unbound cloud environments also differ in their current and recent star formation. The unbound environments have stellar surface density $\Sigma_{star} = 25 - 170\ M_\odot\ pc^{-2}$ while the bound cloud environments have $\Sigma_{star} = 200 - 4000\ M_\odot\ pc^{-2}$ (Schruba et al. 2019; see also Sun et al. 2018, Leroy et al. 2016, Sun et al. 2020).

Thus the prevalence of unbound clumps in most CMZ clouds and of bound clumps in Sgr B2 is not unique to the CMZ of the MW. The CMZ virial properties are especially similar to those in the nuclei of other barred galaxies. Early studies of extragalactic molecular clouds were limited by sensitivity and by selection of Local Group and nearby dwarf galaxies. As more representative star-forming galaxies have been observed, it has become clear that clouds in the nuclei of barred galaxies have significantly greater velocity dispersion and appear less gravitationally bound than their counterparts in the disks of barred galaxies, or in the nuclei of unbarred galaxies (Sun et al. 2018, Sun et al. 2020).

### 6. Clump Mass Distributions (CMDs)

The mass distribution $dN(M)/d\log M$ of a system of star-forming condensations can indicate the relative importance of self-gravity and opposing forces, independent of the foregoing virial analysis, which is restricted to static clumps in equilibrium. The following Section 6.1 reviews CMD models with emphasis on their power-law (PL) slopes. Section 6.2 presents and compares CMDs selected from CMZ populations having a progression of bound clump fractions. Section 6.3 interprets these CMDs with a stopped-accretion model. It derives their virial parameters and relative time scales of accretion and dispersal. It notes that analytic models of competitive accretion and turbulent fragmentation cannot match most of the observed PL slopes.

**6.1. CMD models**

Mass distributions of dense cores in star-forming clouds are often compared to the initial mass function (IMF; Salpeter 1955, Alves et al. 2007, Fiorellino et al. 2021) and to star formation models. The clumps considered here ($M \approx$ 10-300 $M_\odot$) are more massive than dense cores ($M \approx$ 0.3-10 $M_\odot$). Such clumps are often interpreted as progenitors of stellar clusters, while cores are interpreted as progenitors of single stars or small stellar groups. Nonetheless, clump and core mass distributions have each been described as the superposition of two or three different power laws, or as a lognormal function (LN), or as a LN component and a power-law (PL) tail at high mass, with exponent $\Gamma$ (e.g. Brunetti & Wilson 2019; Könyves et al. 2020; O'Neill et al. 2021).

This section describes CMZ clump mass distributions in terms of the LN and PL functions. It also fits CMDs with the LNP function. This function makes a smooth transition from LN to PL shape using a model of stopped accretion (SA; Basu et al. 2015, hereafter BGA15). SA is also known as accretion limited by "equally likely stopping" (Myers 2012) or as "quenched accretion" (Klishin & Chilingarian 2016).

A LN distribution has been ascribed to density fluctuations arising in supersonic MHD turbulence (Padoan et al. 1997), or to multiple stochastic processes in combination with the central limit theorem (Larson 1973, Adams & Fatuzzo 1996). The low-mass portion of the LN component must be interpreted with care, due to possibly incomplete detection of low-mass sources. In this section LN and LNP functions are fit to the observed data, but interpretation of the fits is limited to masses greater than the 95% completeness mass, $M_{\text{comp}} = 50\ M_\odot$ (H20, Section 3.1 and Appendix B).

The PL tail is generally attributed to a significant mass fraction of dense gas which is self-gravitating or collapsing to form stars. Several models have been proposed to explain the PL slope $\Gamma$ of a core mass distribution in the range -1.5 to -1.0, which includes the IMF slope $\Gamma = -1.35$ (Salpeter 1955). However, PL slopes of clump mass distributions need not match the Salpeter slope. The PL slopes of CMDs in the Large Magellanic Cloud are reported to have slopes $\Gamma = -0.9$ to -2.3 (Indebetouw et al. 2013, Brunetti & Wilson 2019).

In models of "competitive accretion" (hereafter CA), a PL slope $\Gamma \approx$ -1 develops if an initial distribution of "seed" masses grows with an accretion rate proportional to $M^2$ as in stationary Bondi accretion (Zinnecker 1982, Bonnell et al. 2007). Numerical simulations yield a mass function with similar slope when protostars in a young cluster compete for accreting gas (Bonnell et al. 2001, Bate 2012, Bate 2014). The slope can become as steep as $\Gamma \approx$ -1.5 if the most massive protostars are concentrated toward the center of the cluster potential (Bonnell et al. 2001).

Turbulent fragmentation (TF) models rely on preferential growth of condensations more massive than the Jeans mass in a turbulent medium (Hennebelle & Chabrier 2008, 2013). An analytic treatment reproduces the IMF with high-mass PL slope approaching $\Gamma = -(1/2)(1 + 3/p)$. Here $p$ is the spectral index of the turbulent velocity spectrum, with expected range $5/3 < p < 2$ (Hopkins 2012; see also Offner et al. 2014 and Krumholz & Federrath 2019). The corresponding range of $\Gamma$ is from -1.25 to -1.4.

In stopped accretion (SA) models, gas accretes onto an initial distribution of seed masses as in CA, but the accretion duration is variable. This duration is the most important factor in setting final mass, according to simulations of protostar formation in clusters (Bate 2012). The duration may be limited by ejection of the accretor from its surrounding gas, or by turbulent motions and protostellar feedback, which reduce the mass of gas available to accrete.

In the simplest SA case, the mass accretion rate is $dm/dt \propto e^{t/\tau_{grow}}$ and the stopping probability density is $p(t) \propto e^{-t/\tau_{stop}}$. The original mass distribution develops a PL tail with slope $\Gamma = -\tau_{grow}/\tau_{stop}$, where $\tau_{grow}$ and $\tau_{stop}$ are characteristic time scale parameters. This SA distribution can describe final masses if the initial masses were born at the same time, or it can represent a snapshot of currently accreting masses, if identical initial masses were born with a uniform birthrate (Basu & Jones 2004, hereafter BJ04; Bate & Bonnell 2005; BGA15).

### 6.2. Comparison of CMDs

The large number of clumps in the CMZ sample, together with the identification of their bound and unbound populations in Sections 3.4 and 3.5, gives a new opportunity to examine how CMDs in the CMZ are related to two indicators of star formation -- the fraction $f_b$ of their bound clumps introduced in section 4.3, and their virial parameter $\alpha$.

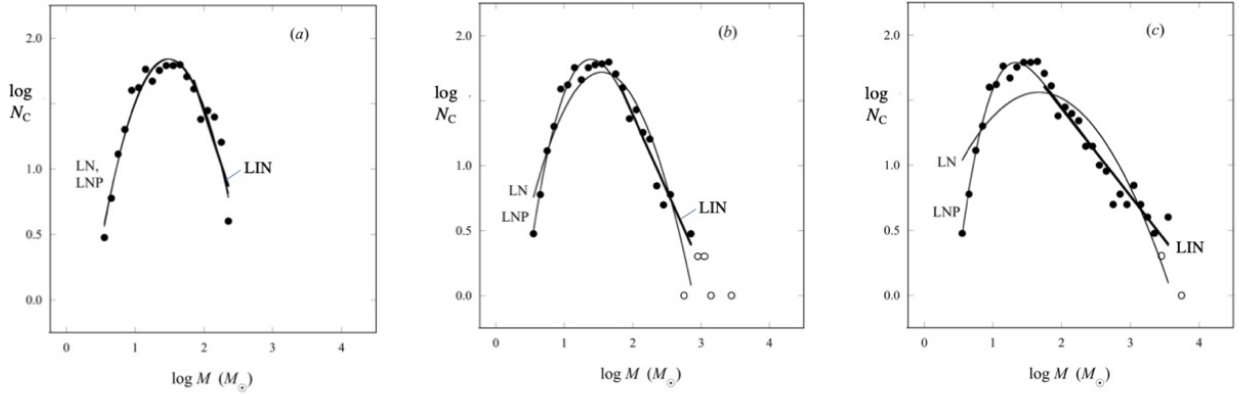

**Figure 7.** Clump mass distributions for CMZ populations with estimated bound clump fraction (a) $f_b = 0$, (b) $f_b = 0.06$, and (c) $f_b = 0.13$. *Filled circles* indicate the log of the number of clumps $N_C$ in a mass bin. The bin width is $\Delta \log(M/M_\odot) = 0.10$, for all bins occupied by at least three clumps. *Open circles* indicate bins occupied by one or two clumps. *Light curves* are LN and LNP model fits to the filled circles. *Heavy lines* are linear fits for clumps more massive than 50 $M_\odot$, the 95% completeness mass. Panel (a) with $f_b = 0$ is for 636 clumps in 22 CMZ clouds, excluding all bound clumps as described in the text. Panel (b) with $f_b = 0.06$ is for 670 clumps in 21 CMZ clouds, excluding Sgr B2. Panel (c) with $f_b = 0.13$ is for all 755 clumps in the 22-cloud CMZ sample, listed in H20 Table 5 and shown in H20 Figure 13. In these plots the slope of the PL component becomes shallower as the bound clump fraction increases.

For this purpose, three CMDs are plotted in Figure 6, with estimated bound clump fractions (a) $f_b = 0$, (b) $f_b = 0.06$, and (c) $f_b = 0.13$. The samples are defined to give a significant progression of $f_b$, taking into account its high estimated value in Sgr B2 and its low value in the remaining 21 CMZ clouds. Sample (b) is considered the most representative of the CMZ since it contains some bound clumps but is not skewed by the exceptionally high number of bound clumps in Sgr B2. The mass bin width is $\Delta \log(M/M_\odot) = 0.10$, which is approximately equal to the one-sigma random uncertainty in $\log(M/M_\odot)$ in Figures 2a-2d. The systematic mass uncertainty is greater than the bin width, but it does not affect the CMD slopes. If each clump mass in a CMD were to change by the same systematic uncertainty, the CMD would shift along the horizontal axis, but its shape and slope would not change. To reduce statistical uncertainty, model fits are applied only to mass bins with three or more clumps. This procedure ensures that each data point used for fitting has an error bar due to Poisson statistics less than or equal to $\sigma_{\log N_C} = 1/(\sqrt{3} \ln 10) = 0.25$.

The distributions are fitted with LN and LNP functions over all allowed bins, and with a linear function over bins more massive than the 95% completeness mass, $M_{\text{comp}} = 50\ M_\odot$ (H20, Section 3.1 and Appendix B). The selection criteria of the CMDs and their linear fit results are summarized in Table 5.

Figure 7(*a*) shows the CMD of 636 clumps in the 22 CMZ clouds in H20 Table 5, after removal of foreground regions, contaminated observations, and all clumps estimated to be bound. These bound clumps are defined to be the most massive clumps in Sgr B2 and in the 21 other CMZ clouds, each according to its bound clump fraction in Sections 4.3 and 5.1. The resulting CMD of unbound clumps in the 22-cloud sample is well-fit by the LN, LNP, and PL functions. The best-fit PL has the steepest slope in Figure 6, with $\Gamma_{u,22} = -1.6 \pm 0.5$. Here the slope subscript $u, 22$ indicates unbound clumps in the 22-cloud sample.

Figure 7(*b*) shows the CMD of the 670 clumps in the 21-cloud population, with $f_{b,21} = 0.06$. This population is chosen to exclude clumps from Sgr B2, to contrast with the distribution in Figure 6(*c*), which includes all clumps. The CMD has a clear PL tail. Its slope is $\Gamma_{b,21} = -1.17 \pm 0.12$, shallower than in Figure 7(*a*). The LNP and LN model fits differ from each other more than the LNP and LN fits in Figure 7(*a*), which are nearly identical.

Figure 7(*c*) shows the CMD of 755 clumps in all 22 clouds, combining clumps in Sgr B2 with those in the 21-cloud population. This figure closely resembles Figure 13 of H20, but with finer mass bins. The mean bound clump fraction is $f_{b,22} = (N_{\text{Sgr B2}} f_{b,\text{Sgr B2}} + N_{21} f_{b,21})/(N_{\text{Sgr B2}} + N_{21}) \approx 0.13$, based on $N_{\text{Sgr B2}} = 85$, $N_{21} = 670$, $f_{b,\text{Sgr B2}} = 0.7$ and $f_{b,21} = 0.06$ estimated in Sections 4.3 and 5.1. The distribution has a prominent power-law tail, with slope $\Gamma_{b,22} = -0.67 \pm 0.05$, the shallowest among the three slopes in Figure 6.

Summarizing, Figure 7 shows that CMDs in the CMZ progress in their shape and PL slope as the bound clump fraction increases. When all clumps are unbound with $f_b = 0$ in Figure 6(*a*), the shape for $M > M_{\text{comp}}$ is nearly that of a LN. This shape coincides with the steepest PL tail, with slope $\Gamma_{u,22} = -1.6$. As the bound fraction increases to $f_{b,21} = 0.06$ in Figure 6(*b*), the PL tail becomes shallower than in Figure 6(*a*), with slope $\Gamma_{b,21} = -1.2$. As the bound fraction increases to $f_{b,22} = 0.13$ in Figure 6(*c*) the PL tail becomes dominant, with the shallowest slope $\Gamma_{b,22} = -0.67$.

A similar change in PL slope is seen in *Herschel* core mass distributions in Orion B. There, the mass distribution of cores having low background is well fit by a log-normal with a PL slope of -1.7 while the distribution of cores in the full sample, which includes more massive cores, has a shallower PL slope -1.3 (Könyves et al. 2020).

### 6.3. Interpretation of CMDs

The CMDs in Figure 7 can be compared with the CA, TF, and SA models discussed in Section 6.1. The closest resemblance appears with SA models of accreting and dispersing clumps. Figure 7($a$), with $f_b \approx 0$, resembles the original LN mass function in Figure 2 of BJ04, when no clumps are yet accreting and when the high-mass slope of the CMD is steeper than at later times. Figures 7($b$) and 7($c$) each resemble the LNP mass function in BJ04, where a PL tail has grown from the original mass function.

In this interpretation the CMDs of low-background cores in Orion B and in Figure 7a have a relatively steep PL slope resembling that of a LN function, with relatively few bound clumps and little star formation. They are considered similar to initial distributions. In contrast, the CMDs in Figure 7b and 7c with shallower PL slopes are considered different possible outcomes from an initial distribution. They differ due to different histories of clump formation and dispersal. However the SA model does not necessarily imply that an ensemble with a given CMD slope will evolve to one with a shallower slope.

This section applies a SA model to infer the accretion rates and virial parameters for the CMDs in Figure 6. To specify the model, it is assumed that the accretion time scale $\tau_{grow}$ is due to self-gravity, and is equal the product of a coefficient $c_{ff}$ and the clump free-fall time $\tau_{ff} = [3\pi/(32G\rho)]^{1/2}$. Here the clump is assumed to be a uniform sphere of density $\rho$. The stopping time scale $\tau_{stop}$ is assumed to be due to clump dispersal, and is equal to the product of a coefficient $c_{cross}$ and the turbulent crossing time $\tau_{cross} = R/\sigma$. The relevance of $\tau_{ff}$, $\tau_{cross}$, and $\tau_{ff}/\tau_{cross}$ to star formation has been discussed by Elmegreen (2000), Padoan et al. (2012), Krumholz (2015), and Schruba et al. (2019).

With the above assumptions, the PL slope can be expressed in terms of the virial parameter $\alpha = 5\sigma^2 R/(GM)$, for the idealized case where the clumps in a CMD have a distribution of masses but the same value of $\alpha$. For clumps following the SA model, $\Gamma = -c_\Gamma \tau_{ff}/\tau_{cross}$ where $c_\Gamma \equiv c_{ff}/c_{cross}$. Then $\tau_{ff}/\tau_{cross}$ can be expressed in terms of $\alpha$ (Schruba et al. 2019), yielding

$$\Gamma = -(\pi c_\Gamma/2)(\alpha/10)^{1/2}. \qquad (15)$$

Equation (15) gives a simple connection between the PL slope of a CMD and a virial parameter $\alpha$ typical of its clumps. Evaluating $\alpha$ for the three best-fit linear slopes in Figure 7 indicates highly unbound clumps in (*a*), moderately unbound clumps in (*b*), and bound clumps in (*c*), provided the coefficient $c_\Gamma$ is close to unity in each case.

The selected clump populations in Figure 7 differ only in their numbers of clumps more massive than $\sim 100\ M_\odot$. Thus their values of $\alpha$ derived from equation (15) indicate the degree of gravitational binding only among these most massive clumps, which also set the slope $\Gamma$. The properties of the three CMD populations, their PL slopes, and their derived virial parameters are summarized in Table 5.

**Table 5**

CMD Properties and SA Model Parameters

| (1) Selection | (2) Clouds | (3) Clumps | (4) $f_b$ | (5) $\Gamma = -\tau_{grow}/\tau_{stop}$ | (6) $\alpha c_\Gamma^2$ |
|---|---|---|---|---|---|
| (*a*) all unbound clumps | 22 | 636 | 0 | $-1.6 \pm 0.5$ | $9 \pm 6$ |
| (*b*) all clumps not in Sgr B2 | 21 | 670 | 0.06 | $-1.2 \pm 0.1$ | $6 \pm 1$ |
| (*c*) all clumps | 22 | 755 | 0.13 | $-0.67 \pm 0.05$ | $1.8 \pm 0.3$ |

**Note**- Column (1) gives selection criteria for the clump mass distributions (CMDs) in Figures 6(*a*), 6(*b*), and 6(*c*). Columns (2), (3), and (4) give respectively the number of clouds, number of clumps, and the fraction of bound clumps $f_b$ in each CMD. Column (5) gives the best-fit linear slope and its one-sigma uncertainty, for CMD clumps in the power-law (PL) mass range, and the corresponding time scale ratio in the SA model. The fitting is based on the Levenberg-Marquardt algorithm (Levenberg 1944; Marquardt 1963). Column (6) gives the product of the virial parameter $\alpha$ in the PL mass range derived from equation (15), and the coefficient $c_\Gamma^2$. This coefficient is the squared ratio of the clump accretion time scale normalized by the free-fall time to the clump dispersal time scale normalized by the turbulent crossing time. The uncertainties in column (6) are propagated from those in column (5).

Table 5 shows that the SA model matches the PL slope of each CMD in Figure 6 to a ratio of accretion and dispersal time scales and to a characteristic virial parameter. In case (*a*) the CMD with few if any bound clumps has the highest virial parameter, and its clumps tend to disperse faster than they can accrete, by a factor ~2. In contrast, in case (*c*) the CMD with the most bound clumps has the lowest virial parameter, its clumps tend to accrete faster than they can disperse, by a factor ~1.5, and it is associated with numerous young clusters and massive protostars. The most representative CMD in case (*b*) is also intermediate in its ratio of time scales, virial parameter, and degree of star formation.

The time scale ratios in Table 5 can be interpreted in terms of clump free-fall time, crossing time, and the individual time scales. For the representative case (*b*) the PL slope is $\Gamma = -1.2$, the typical virial parameter is $\alpha = 6/c_\Gamma^2$ and the 670 values of clump density give mean free-fall time $\tau_{ff} = 0.076$ Myr. Assuming that this value of $\alpha$ equals the mean value $\alpha = 10$ of the virial model fits in Table 4 gives $c_\Gamma \approx (3/5)^{1/2} = 0.77$. Then the typical turbulent clump crossing time is $\tau_{cross} = (-\tau_{ff} c_\Gamma)/\Gamma = 0.049$ Myr. The corresponding growth and stopping time scales can each be written as a small multiple of the free-fall time, $\tau_{grow}/\tau_{ff} = c_{ff}$ and $\tau_{stop}/\tau_{ff} = -c_{ff}/\Gamma$. As an example if $c_{ff} = 3$, $\tau_{grow} = 0.23$ Myr and $\tau_{stop} = 0.19$ Myr. The similarity of these values is consistent with the idea that the typical unbound clump cannot grow significantly before it disperses.

This application of the SA model does not specify the clump birthrate and initial masses, and it does not justify the assumed exponential form of the mass accretion rate. These features are needed for a more complete treatment. Nonetheless the SA predictions of reasonable time scale ratios and virial parameters motivate more detailed applications in future studies.

In contrast to the SA results, it is more difficult to explain the variation of slopes in Figure 7 with the CA or TF models discussed in Section 6.1. To explain the slopes $\Gamma = -0.67$ and $\Gamma = -1.2$ with the TF model of Hopkins (2012) would require $2.1 \leq p \leq 8.8$, values of the turbulent velocity spectral index which lie outside the expected range $5/3 \leq p \leq 2$. Similarly, the slopes $\Gamma = -0.67$ and $\Gamma = -1.6$ lie outside the range of slopes -1 in the CA model of Zinnecker (1982) to -1.5 in the CA model of Bonnell et al. (2001). It remains to test these and other models with more detailed observational and numerical studies.

# 7. Discussion

This section summarizes the paper, describes its limitations and uncertainties, and discusses how the new findings of the paper relate to the suppression of star formation in the CMZ.

## 7.1. Summary

This paper investigates the star-forming potential of 755 clumps in 22 clouds in the Central Molecular Zone of the Milky Way, in order to better understand the relatively low star formation rate in the CMZ. The paper analyzes observations of dust emission at 1.3 mm wavelength by the Submillimeter Array in the CMZoom project (B20, H20). These are the most extensive continuum observations available of the dense gas in the CMZ, with resolution ~3" or 0.1 pc. The observations are sensitive to clump column densities greater than ~ $10^{23}$ cm$^{-2}$ and masses greater than ~10 $M_\odot$. They have a 95% completeness mass of 50 $M_\odot$.

The observations are analyzed with a model of virial equilibrium which includes effects of self-gravity, external pressure, and magnetic fields, for uniform spherical condensations in the mass range of "clumps" (S78, F11, Bergin & Tafalla 2007). Standard virial analysis of CMZoom observations requires well-determined velocity dispersions, which are available for about 0.1 of the observed clumps (Walker et al. 2018, Callanan 2021). To analyze the observations in more detail, this work uses a modified form of virial analysis, which is applied to the entire CMZoom catalog (H20 Table 5).

In "$N - M$ virial analysis" the observed clump column densities $N$ and masses $M$ are plotted in the $\log N - \log M$ plane. Virial model curves are superposed, for varying values of velocity dispersion $\sigma$. In this method $\sigma$ is a parameter rather than an observable as in standard virial analysis. For unbound virial clumps in regions of similar external pressure, $N \propto M^s$, where $s \gtrsim 1/3$, and where $N$ is only weakly dependent on $\sigma$. Thus $N - M$ virial analysis is especially useful for large samples having a high fraction of unbound clumps and a low fraction of available velocity dispersion measurements. Section 2 describes this form of virial analysis and compares its features with those of standard virial analysis.

Section 3 presents the observations of 24 CMZoom clouds, in the form of 24 $\log N - \log M$ plots, each with the same template of nine virial curves. The curves vary in their values of velocity dispersion and critical column density. The plots are presented in four groups of six, ordered by increasing clump mass and by the complexity of the distribution. Nearly all plots show an approximately linear trend $\log N \propto \log M$ with slope $s \gtrsim 1/3$, in the mass range $10 \lesssim M \lesssim 200\ M_\odot$. In contrast, Sgr B2 shows a more complex distribution of clumps having significantly greater values of $\log N$ and $\log M$.

Section 4 analyzes the linear trends in the plots in Section 3. The trend in each cloud plot is fitted with a linear function to estimate its slope and intercept. The distribution of these 22 "cloud slopes" has mean ± standard error $0.38 \pm 0.03$. As a check against possible bias a uniformly weighted distribution of 595 "clump slopes" was also computed. This distribution of slopes has essentially the same mean and uncertainty as the first. Together the distributions indicate a slope only slightly greater than the slope $s_p = 1/3$ in the pressure-confined limit of virial equilibrium. This result suggests that most CMZ clouds have a large population of gravitationally unbound clumps.

Nine of the clouds with linear $\log N - \log M$ trends have one or two clumps with significantly greater values of $\log N$ and $\log M$. Their values of $\log N$ are greater than expected from the linear trend, but are consistent with the part of a virial curve where the pressure branch approaches the self-gravitating branch. Virial model fits to these "bound-and-unbound" systems give estimates of the critical column density, velocity dispersion, external pressure, virial parameter, and bound clump fraction. The nine-cloud average of the median virial parameters is $\alpha = 10$. This significantly unbound value has corresponding $\log N - \log M$ slope $s = 0.38$ according to equation (9), consistent with the linear fit slope histograms in Figure 3 and 4 and supporting their unbound interpretation.

The velocity dispersions, virial parameters, and bound clump fractions derived from these bound-and-unbound fits are in good agreement with independent estimates in several CMZ clouds observed with finer resolution.

Section 5 estimates the bound clump fraction in Sgr B2 to be $0.7 \pm 0.1$, based on association of SMA clump positions with ALMA 3 mm sources, which are interpreted to be high-mass protostellar objects or protostellar clusters (G18). This high fraction contrasts with the typical bound clump fraction 0.06 in the nine clouds analyzed in Section 4. The massive clumps in Sgr B2 show a linear trend steeper than that of unbound low-mass clumps. Its slope $s \approx 0.5$ is consistent with critically bound clumps whose velocity dispersion increases with mass. The differing properties between Sgr B2 and most of the other CMZ clouds resembles the two types of virial equilibrium systems identified in studies of MW clouds and of clouds in barred galaxies (Schruba et al. 2019, Sun et al. 2020).

Section 6 compares the shapes of clump mass distributions (CMDs) selected from the full sample of CMZ clumps to have a progression of bound clump fractions $f_b \approx 0, 0.06$, and $0.13$. Their power-law slopes $\Gamma$ progress from steep $(\Gamma = -1.6)$ to shallow $(\Gamma = -0.67)$, in agreement with a model of stopped accretion. In this model $\Gamma$ increases with the virial parameter and is proportional to the ratio of the rates of clump dispersal and accretion. For the typical CMD $(\Gamma = -1.2)$, the typical clump is unbound and has rates of dispersal and accretion of order 0.2 Myr. This model fits the observed values of $\Gamma$ better than do analytic models of competitive accretion or turbulent fragmentation.

## 7.2. Limitations and Uncertainties
### 7.2.1. Limitations

The virial model of CMZ clumps in Sections 2-5 incorporates effects of self-gravity, external pressure, and magnetic fields, but it is nonetheless an equilibrium model of static clumps. It does not address the dynamical questions of how clumps form, grow, collapse, and disperse. The relative rates of clump accretion and dispersal are estimated in a highly simplified way in the SA analysis of the three CMDs in Section 6. Each of these approaches gives a partial view of a more complex picture. Simulations of the evolution of turbulent clouds which track the binding of individual clumps may be useful for comparison.

The $N - M$ virial model introduced in Section 2 identified virial systems of weakly bound clumps without exact knowledge of their velocity dispersions. This result is expected when the variations in $n \propto P_S/\sigma^2$ from clump to clump are small enough to allow the trend $N \propto M^p, p \gtrsim 1/3$ to be recognized, as noted in Section 2.2.1. Nonetheless it will be important to extend the work of Walker et al. (2018) and Callanan (2021) to obtain observed velocity dispersions for the full clump sample, to compare bound clump fractions for individual clouds with those estimated here. Well-determined velocity dispersions are especially important to identify clumps which are nearly bound.

Perhaps the most useful procedure to estimate clump binding and to test the $N - M$ virial model is to apply both the standard and the $N - M$ methods of PVE analysis to the same set of bound and unbound clumps, provided the set is large enough to reveal the statistical trends analyzed here. This "training set" experiment would be useful to reveal the areas of agreement and disagreement between the methods.

The radial profiles of dust clumps may vary with their evolutionary stage, since more evolved sources tend to have a steeper intensity profile (e.g. Beuther et al. 2002). This property may affect the dendrogram-measured clump size. It is also likely that sufficiently compact and hot evolved star-forming sources within the CMZoom sample require higher resolution measurements of their temperature structure to accurately model their physical properties. In this case, we would expect gravitationally bound sources to have higher temperatures (and therefore lower masses) than those listed in the catalog. In terms of the N-M model, this would likely mean the high mass points might be biased to higher mass and column densities. However, these points are already excluded in the intercept and slope fits described in section 4.1, so this possible bias would not affect the derived slopes in Figures 3 and 4.

Molecular line studies are also needed to determine how much the velocity dispersion at a given clump size scale varies from one line to the next, and from one size scale to the next. On the one hand, ALMA line observations of clumps in the 50 km s$^{-1}$ cloud show substantial consistency between velocity dispersions ~2 km s$^{-1}$ in the lines of C$^{34}$S 2-1 and H$^{13}$CO$^+$ 1-0 (Uehara et al. 2019). On the other hand, analysis of emission in G0.253+0.016 (the "Brick") in ALMA line observations of HNCO 4(0,4)-3(0.3) shows spatially complex velocity structure with evidence of large-scale oscillations and shear (Henshaw et al. 2019).

The virial analysis in Sections 3 and 4 indicates that most clumps are close to pressure balance with their turbulent environmental gas. This key result may be questioned if the turbulent gas has large-amplitude pressure fluctuations, in which case a weakly bound clump may be transient, and dispersed in a few crossing times. It may be possible to test this point with observations of gas density and velocity dispersion inside and outside clump boundaries, in spectral lines which trace high and low density gas. A similar procedure was carried out for dense cores in Orion A (Kirk et al. 2017).

The properties derived from virial analysis in Section 4 depend only weakly on the assumed mass-to-magnetic-flux ratio $\lambda$, as long as the clumps are at least slightly magnetically supercritical. The property calculations in Section 4 assume a negligibly weak field, with $\lambda = \infty$ or $c_M = 1$. For stronger fields, the virial parameter $\alpha$ varies as $c_M = 1 - \lambda^{-2}$ for weakly bound clumps and the velocity dispersion $\sigma$ varies as $c_M^{1/2}$. Then $\alpha$ will be reduced by a factor less than 2 as long as $\lambda > \sqrt{2} = 1.41$ and $\sigma$ will be reduced by a factor less than 2 as long as $\lambda > 2/\sqrt{3} = 1.16$. These limiting mass-to-flux ratios are similar to estimates of $\lambda$ in star-forming dense cores (Myers &

Basu 2021). It will be important to estimate magnetic field strengths and mass-to-flux ratios in CMZ clouds to better constrain the role of the magnetic field in the foregoing virial analysis.

The values of bound clump fraction $f_b$ estimated in sections 4.3 and 5.1 are important indicators of the star formation potential of a CMZ cloud, but at present they are too poorly known to provide a quantitative ranking of $f_b$ in CMZ clouds. Such a ranking would be valuable to compare to the prevalence of tracers of recent star formation such as H II regions, CO outflows, and $H_2O$ and $CH_3OH$ masers in the same set of clouds. The current precision of $f_b$ can be greatly improved when sensitive, well-resolved velocity dispersions become available for the full sample of CMZoom clumps.

### 7.2.2. Quantitative Uncertainties

As mentioned in Section 3.1, the measurements of column density $N$ and mass $M$ are dominated by systematic uncertainties due to the assumptions about dust opacity, temperature, and dust-to-gas ratio used in their calculation. However, changes in $\kappa_\nu$, $T$ and $R_{dg}$ scale identically with $N$ and $M$. Thus uncertainties in slopes $s = (M/N)(dN/dM)$, radii $R \propto (M/N)^{1/2}$, and velocity dispersions $\sigma \propto (M/N)^{1/4}$ in Section 5.2 should be due only to the random components of the uncertainties in $N$ and $M$. In contrast, clump densities $n \propto N(N/M)^{1/2}$ have both systematic and random uncertainties. The critical column density $N_0$ derived from model fits in Section 4.3 has random uncertainty comparable to its fit uncertainty, but its systematic uncertainty is probably similar to that of $N$. Thus the uncertainty in $P_S \propto N_0^2$ is likely dominated by the systematic uncertainty in $N$.

The uncertainties in $N$ and $M$ are calculated using equations (10) and (11). The random uncertainties vary from leaf to leaf within a cloud, while the systematic uncertainties are unlikely to vary significantly within a cloud. The random errors are propagated through equations (10) and (11) from the uncertainties in local 1.3 mm dust continuum flux and local dust temperature fluctuations. The local noise properties and the typical expected temperature fluctuations are discussed in section 4.3 of H20 in more detail. These random uncertainties impact the results of all parameters fit from the mass and column densities of CMZoom objects. The systematic uncertainties, which tend to dominate the total uncertainties for the dust continuum catalogs, are estimated from our assumptions about the gas-to-dust ratio, dust opacity, and distance to the individual cloud within the Galactic Center. The expected uncertainties in gas-to-dust ratio and dust opacity are largely unconstrained within the Galactic Center, so we adopt a factor of two for their systematic uncertainty, as suggested by Battersby et al. (2010). The uncertainty in line-of-sight distance is chosen to be ~240 pc, corresponding to the maximum longitudinal extent of the CMZoom source sample.

Following the above procedures, the average relative random uncertainties for the leaf masses and column densities for all leaves in the catalog are $(\sigma_M/M)_{\text{ran}} = 0.23 \pm 0.04$ and $(\sigma_N/N)_{\text{ran}} = 0.32 \pm 0.10$. These values take into account the uncertainties due to distance, dust temperature, and local noise. The total relative uncertainties, including systematic errors, are dominated by the gas-to-dust ratio and dust opacity. They are $(\sigma_M/M)_{\text{tot}} = 2.010 \pm 0.004$ and $(\sigma_N/N)_{\text{tot}} = 2.030 \pm 0.015$.

### 7.3. Spatial Structure in CMZ Clouds

The clump model used in this paper is unrealistic in its spatial detail because it assumes spherical clumps in a uniform medium. In contrast CMZ observations show spatial structures which resemble nearby star-forming regions, where many clumps have elongated contours of intensity. They are organized into filamentary networks which often have a few massive clumps in central locations, and numerous lower-mass clumps embedded in extended filaments (e.g. Lu et al. 2020, 2021; H20). Clumps are part of a complex hierarchy of structure on many scales, and the resolution of the CMZoom observations limits conclusions about CMZ clumps to scales $\gtrsim 0.1$ pc. Thus it is expected that many CMZ clumps in the H20 catalog will reveal more complex substructure at finer resolution, like that reported by Lu et al. (2020) and Lu et al. (2021).

These observations of the filamentary structure of the CMZ clump environment do not necessarily invalidate the virial analysis presented here and elsewhere (Walker et al. 2018, Callanan 2021). However they suggest that more realistic models should reflect the anisotropic gravitational force and pressure exerted on a clump by its filamentary environment.

### 7.4. Suppression of Star Formation in the CMZ

This paper presents new evidence that clumps in nearly all CMZ clouds are gravitationally unbound, based on a more extensive and complete survey (the CMZoom survey) of CMZ clouds than was previously available. These new findings include (1) Linear trends of $\log N$ with $\log M$ for ~20 clumps in each of in 22 clouds. These trends are parallel to the pressure branch of the virial curve, and their slope is close to the pressure-bound limit $s_0 = 1/3$. (2) Nine clouds with both a linear trend and one or two more massive clumps are fit by virial models of both bound and unbound clumps. These models indicate a population dominated by unbound clumps, with typical virial parameter ≈ 10. (3) In contrast, Sgr B2 shows trends consistent with a large population of bound clumps and small population of unbound clumps. (4) Clump mass distributions (CMDs) have power-law (PL) slopes fit by a model of accreting and dispersing clumps. The typical PL slope implies that the typical clump is unbound, and is unlikely to accrete significantly before it is dispersed.

These new findings provide a clear link to the suppressed star formation in the CMZ discussed in Section 1.2, on the premise that gravitationally unbound clumps are very unlikely to form stars. It is now clear that low bound clump fraction is a widespread property of nearly all the clouds in the CMZ. They are unbound not because their density is unusually low, but rather because their turbulence is unusually high, probably due to flows from the galactic bar into the CMZ (Sormani et al. 2020, Hatchfield et al. 2021), or to tidal compression during percenter passage (Kruijssen et al. 2019, Dale et al. 2019). The model of inflow from the galactic bar is supported by observations which find median velocity dispersion in the nuclei of 43 barred galaxies, greater than in their galactic disks and in the nuclei of unbarred galaxies, each by a factor ~5 (Sun et al. 2020).

The widespread evidence for low bound clump fraction in the CMZ could also be consistent with quiescent star formation as the low part of a cycle of episodic star formation (Krumholz et al. 2015, Armilotta et al. 2019), discussed in Section 1.2. However, such cycling might be inconsistent with the observations of velocity dispersion in galactic nuclei cited above. If most nuclear clouds in barred galaxies cycled through quiescent and starburst phases, their cycles would be expected to have random phases from one galaxy to the next. If the high and low part of each cycle had similar duration, a sample of ~40 galaxies might have a distribution of high and low velocity dispersions rather than the observed concentration of high values in Figure 2 of Sun et al. (2020).

As noted in Section 1.2, any explanation of a quiescent CMZ must also explain the starburst properties of Sgr B2. Such starburst clouds may arise by collision of gas flowing from the bar to the CMZ with already orbiting CMZ gas (Sormani et al. 2020). It may therefore be useful to search for analogs of Sgr B2 in the quiescent CMZs of other barred galaxies, to better understand their origin.

## 8. Conclusions

This paper presents results of the largest complete survey of high-column-density gas in the CMZ of the Milky Way, based on the column densities $N$ and masses $M$ of 755 clumps in 22 clouds (B20, H20). The data are presented in the $\log N - \log M$ plane for each cloud. They are analyzed with a virial equilibrium model including self-gravity, magnetic fields, and external pressure. The model is formulated in the $\log N - \log M$ plane for comparison with the data.

The main result is that in 21 CMZ clouds (all except Sgr B2), gravitationally bound clumps are rare, with bound fraction $f_b \lesssim 0.06$. This is the most extensive estimate of CMZ clump binding available. It offers the clearest basis to date for suppressed star formation in the CMZ. The following are the main details:

1. In 22 clouds, 595 low-mass clumps show approximately linear trends $\log N \propto s \log M$ with typical slope $s = 0.38 \pm 0.03$. These trends coincide with the pressure branch of a virial curve. Their slope is close to the pressure-bound limit $s_p = 1/3$, indicating a dominant population of gravitationally unbound clumps.

2. In nine clouds, clumps with a linear trend are associated with one or two distinctly more massive clumps. Virial model fits which assume that the most massive clump is critically bound give estimates of the critical column density, virial parameter, density, velocity dispersion, and bound clump fraction for more than 200 clumps. The mean virial parameter is $\alpha = 10$ and the mean bound clump fraction is 0.06.

3. The bound clump fraction in Sgr B2 is close to 0.7, based on association with 3mm ALMA sources (Ginsburg 2018). This fraction is ~10 times greater than in all other CMZ clouds. Sgr B2 clumps have two significant linear trends. Ten low-mass clumps with slope 0.3 resemble the unbound clumps in other CMZ clouds. Seventy-three massive clumps with slope 0.5 are consistent with critical binding where the velocity dispersion increases with mass. Such clouds dominated by bound or unbound clumps resemble the two types of VE identified in other MW clouds, and in nuclei of barred and unbarred galaxies.

4. Clump mass distributions (CMDs) in the CMZ are selected to study their change of shape with increasing bound clump fraction $f_b$. Their power-law (PL) slopes become shallower with increasing $f_b$. Their range of slopes is consistent with a model of stopped accretion (SA), but lies outside the range of slopes in models of competitive accretion and turbulent fragmentation.

5. A SA model of the CMD which has typical PL slope indicates an unbound virial parameter similar to that from virial studies, and similar time scales of clump growth and dispersal, of order 0.1 Myr. This similarity implies that the typical unbound clump cannot gain significant mass before the typical time of its dispersal.

6. The prevalence of unbound clumps in CMZ clouds naturally accounts for suppressed star formation in the CMZ, on the premise that stars and clusters can form with much greater efficiency from bound clumps than from unbound clumps. The prevalence of unbound clumps is probably due to high levels of turbulence in the CMZ, since elevated velocity dispersions are an observed property of nuclei of barred galaxies. The high turbulence may be driven by inflow from the galactic bar to the CMZ.

**Acknowledgements.** We acknowledge the Scientific Editor of this paper, Judith Pipher, for her exceptional service to the astronomical community over her distinguished career. We thank the referee for a helpful report which improved the paper. PCM thanks the organizers of the New England Regional Star Formation Meeting at the University of Connecticut in 2020, where this work was presented in preliminary form. PCM also thanks Jens Kauffmann, Thushara Pillai, Mike Dunham, and Sarah Sadavoy for helpful discussions, and Terry Marshall for support. H P.H. gratefully acknowledges support from the SOFIA Archival Research Program (program ID 09_0540). H P.H. also thanks the LSSTC Data Science Fellowship Program, which is funded by LSSTC, NSF Cybertraining grant #1829740, the Brinson Foundation, and the Moore Foundation; his participation in the program has benefited this work. H P.H. gratefully acknowledges support from the National Science Foundation under Award No. 1816715. CB gratefully acknowledges support from the National Science Foundation under Award Nos. 2108938 and 1816715 and from the SOFIA Archival Research Program (program ID 09_0540). We wish to recognize and acknowledge the very significant cultural role and reverence that the summit of Maunakea has always had within the indigenous Hawaiian community. We are most fortunate to have had the opportunity to use observations from this mountain.